\documentclass[a4paper,fleqn]{cas-sc}

\usepackage[authoryear,longnamesfirst]{natbib}
\usepackage{graphicx}
\usepackage[export]{adjustbox}

\usepackage{subcaption}

\def\tsc#1{\csdef{#1}{\textsc{\lowercase{#1}}\xspace}}
\tsc{WGM}
\tsc{QE}
\tsc{EP}
\tsc{PMS}
\tsc{BEC}
\tsc{DE}
\begin{document}
\let\WriteBookmarks\relax
\def\floatpagepagefraction{1}
\def\textpagefraction{.001}

% Short title
\shorttitle{Additive Manufacturing of PEEK/Lunar Regolith ...}

% Short author
\shortauthors{Mohammad Azami et~al.}

% Main title of the paper
\title [mode = title]{Additive Manufacturing of PEEK/Lunar Regolith Composites for Sustainable Lunar Manufacturing}                      
% Title footnote mark
% eg: \tnotemark[1]
%\tnotemark[1,2]

% Title footnote 1.
% eg: \tnotetext[1]{Title footnote text}
% \tnotetext[<tnote number>]{<tnote text>} 
%\tnotetext[1]{This document is the results of the research
 %  project funded by the National Science Foundation.}

%\tnotetext[2]{The second title footnote which is a longer text matter
 %  to fill through the whole text width and overflow into
  % another line in the footnotes area of the first page.}

% First author
%
% Options: Use if required
% eg: \author[1,3]{Author Name}[type=editor,
%       style=chinese,
%       auid=000,
%       bioid=1,
%       prefix=Sir,
%       orcid=0000-0000-0000-0000,
%       facebook=<facebook id>,
%       twitter=<twitter id>,
%       linkedin=<linkedin id>,
%       gplus=<gplus id>]
\author[1]{Mohammad Azami}[type=editor,
orcid=0000-0002-7086-8994]

% Corresponding author indication
%\cormark[1]

% Footnote of the first author
%\fnmark[1]

% Email id of the first author
\ead{mohammad.azami@mail.concordia.ca}

% URL of the first author
%\ead[url]{www.cvr.cc, cvr@sayahna.org}

%  Credit authorship
%\credit{Conceptualization of this study, Methodology, Software}

% Address/affiliation
\affiliation[1]{organization={Department of Electrical and Computer Engineering, Concordia University},
  %  addressline={Radarweg 29}, 
city={Montréal},
province={Québec,},
    % citysep={}, % Uncomment if no comma needed between city and postcode
   % postcode={1043 NX}, 
    % state={},
country={Canada}}

% Second author
\author[1]{Pierre-Lucas Aubin-Fournier}

\ead{pierrelucas.aubinfournier@concordia.ca}

\author [2] {Mehdi Hojjati}[type=editor,
orcid=0000-0002-4674-9397]

%\fnmark[1,3]
\ead{mehdi.hojjati@concordia.ca}
\ead[URL]{https://www.concordia.ca/faculty/mehdi-hojjati.html}
\affiliation[2]{organization={Department of Mechanical, Industrial, and Aerospace Engineering, Concordia University},
  %  addressline={Radarweg 29}, 
city={Montréal},
province={Québec,},
country={Canada}}

    % Corresponding author
\author [1] {Krzysztof Skonieczny}[type=editor,
orcid=0000-0002-6540-3922]
\cormark[1]
%\fnmark[1,3]
\ead{krzysztof.skonieczny@concordia.ca}
\ead[URL]{https://www.concordia.ca/faculty/krzysztof-skonieczny.html}

% Corresponding author text
\cortext[cor1]{Corresponding author}
%\cortext[cor1]{Corresponding author}

% Here goes the abstract

\begin{abstract}
As humanity advances toward long-term lunar presence under NASA’s \emph{Artemis} program, the development of lunar-based manufacturing and construction (LBMC) capabilities has become increasingly critical. The high cost of transporting materials from Earth makes \emph{in-situ resource utilization} (ISRU) essential, with lunar regolith serving as a promising local feedstock. Additive manufacturing (AM) offers a compelling platform for LBMC due to its geometric flexibility, material efficiency, and capacity for on-demand, site-specific production. This study investigates material extrusion (MEX) AM of polyether–ether–ketone (PEEK) composites containing 10–50~wt\% lunar regolith simulant (LRS). PEEK and LMS-1D powders were melt-compounded via twin-screw extrusion, printed using a high-temperature chamber, and annealed at 300~$^{\circ}$C. The samples were characterized through density measurements, thermal analysis, tensile testing, and microstructural and elemental mapping. All filaments exhibited densities above 96~\%, though as-printed porosity increased from less than 1~\% in neat PEEK to 7.5~\% at 50~wt\% LRS due to elevated melt viscosity. Regolith incorporation enhanced crystallinity (17.4 to 20.5\%) and elastic modulus (by 6--41\%), while reducing delamination and warping, which improved dimensional accuracy and print success rates. Tensile strength declined gradually from 107~MPa to 90~MPa up to 40~wt\% LRS, then dropped sharply to $\sim$70~MPa at 50~wt\%. Annealing improved density and stiffness for composites containing up to 30~wt\% LRS, with marginal benefit at higher contents. Microstructural and elemental analyses confirmed a continuous PEEK matrix with uniformly dispersed regolith particles. This work establishes processing windows and trade-offs for regolith-rich PEEK composites, supporting ISRU-enabled AM of future lunar infrastructure.

\end{abstract}

% Use if graphical abstract is present
% \begin{graphicalabstract}
% \includegraphics{figs/grabs.pdf}
% \end{graphicalabstract}

% Research highlights
\begin{highlights}

    \item Successfully demonstrated material extrusion (MEX) additive manufacturing of PEEK composites containing up to 50~wt\% lunar regolith simulant, which is the highest loading reported to date.

    \item Regolith addition reduced delamination and warping, leading to improved dimensional accuracy and higher print success rates compared to pure PEEK.
    
    \item Regolith incorporation increased the elastic modulus by 6--41\% and enhanced matrix crystallinity by approximately 3 percentage points.

    \item Post-print annealing improved density and crystallinity for blends up to 30~wt\% LRS, with limited benefit observed at higher loadings.
    
    \item Microstructural analysis confirmed well-dispersed regolith particles and a distinct contrast between the carbon-rich matrix and oxide-based filler.

\end{highlights}

% Keywords
% Each keyword is separated by \sep

\maketitle

\section{Introduction}

NASA’s Artemis program marks the beginning of a new era of sustained human exploration on the Moon, requiring the development of substantial infrastructure to support long-term human presence and mission activities on the lunar surface~\cite{gelino2024selection}. Developing all required materials and components on Earth and subsequently transporting them to the lunar surface is neither practical nor sustainable, especially considering significant mass and volume constraints. Furthermore, if a component or structure on the lunar surface requires development, replacement, or repair, it is impractical and unsafe for astronauts to await deliveries from Earth. Consequently, in-space manufacturing (ISM), and more specifically lunar-based manufacturing and construction (LBMC), has become a central element in the strategic planning of future lunar missions, regardless of the organization involved.~\cite{azami2024comprehensive, isachenkov2021regolith, zocca2022challenges, altun2021additive}.

Significant volume and mass launch costs and limitations underscore the critical need for maximizing the utilization of the Moon's local resources through in-situ resource utilization (ISRU). Therefore, reducing dependence on Earth-sourced raw materials by leveraging available lunar regolith and other indigenous resources is essential~\cite{araghi2022nasa, sanders2025progress}.

To enhance operational efficiency and reduce dependency on heavy equipment transported from Earth, additive manufacturing (AM) techniques offer substantial advantages. These methods are preferred due to their minimal material waste, reduced energy consumption, increased geometric freedom, and enhanced customization capabilities~\cite{azami2023laser, abedi2019high, kazemi2025uncertainty, kazemi2022uncertainty}. Such capabilities are particularly valuable since many lunar-manufactured components are likely to be unique or unprecedented. Therefore, additive manufacturing, using compact printers for smaller parts and mobile, scalable systems for larger structural elements, represents a critical technological pathway toward sustainable, efficient, and responsive lunar manufacturing and infrastructure development~\cite{hoffmann2023space, azami2024comprehensive, li2020slaam}.

Incorporation of lunar regolith into additive manufacturing is thus an active area of research \cite{azami2024comprehensive, isachenkov2021regolith}. The limited availability of authentic lunar regolith on Earth, brought back by lunar missions, poses considerable challenges for technology development and validation for future lunar applications. To address this limitation, several international research groups have developed lunar regolith simulants (LRS), which are synthetic materials engineered to replicate the physical and chemical properties of lunar regolith~\cite{li2022preparation, mueller2014additive, creedon2023development, wang2022situ}.

Ensuring reproducibility on the Moon of testing results with simulants motivates manufacturing processes that exhibit minimal sensitivity to variations in regolith geochemistry. This requirement arises from two primary considerations. Firstly, current simulants cannot fully replicate all lunar regolith properties because accurately reproducing lunar conditions on Earth is inherently difficult~\cite{taylor2016evaluations, iantaffi2025laser}. Lunar regolith comprises a diverse mixture of highly angular and irregular mineral fragments, agglutinates, dust particles ($<20\,\mu\mathrm{m}$), small rocks ($<1\,\mathrm{cm}$), and larger boulders, varying significantly in composition, cohesion, packing density, abrasion resistance, and flow characteristics.
These variations result from regolith maturity, location-specific geochemical processes, and environmental exposure to meteorite impacts, solar wind, and galactic cosmic radiation~\cite{iantaffi2025laser}. Secondly, significant variability in lunar regolith properties across different lunar regions further complicates direct replication efforts~\cite{heiken1991lunar}.

Consequently, binder-based manufacturing methods are highly attractive for lunar applications due to their reduced sensitivity to variations in regolith composition, especially when compared to sintering or melting-based techniques such as powder-bed fusion. Among these binder-based methods, material extrusion (MEX), specifically fused filament fabrication (FFF), stands out as particularly advantageous due to the use of solid-state thermoplastic binders~\cite{azami2025enhancing}. This preference arises because traditional liquid binders face substantial operational challenges on the lunar surface, including incompatibility with NASA’s stringent outgassing regulations in vacuum conditions and risks of boiling or freezing under the Moon’s extreme temperature fluctuations. Furthermore, FFF-based material extrusion offers additional benefits such as process simplicity, minimal reliance on large, heavy equipment, and demonstrated efficiency in low-gravity environments \cite{prater2016summary, werkheiser2015space, sacco2019additive, huang2025experimental}. Notably, MEX is highly compatible with robotic systems, particularly small mobile robots, which are preferred for ISM due to the critical need to minimize the mass and volume of equipment \cite{li2020slaam}. This makes MEX a promising approach for the construction of large-scale structures and components in future lunar and space missions.

Among the thermoplastic binders used for MEX, many are not suitable for the harsh thermal conditions of the lunar surface, particularly during the lunar day when temperatures can reach $120$--$130\,^\circ\mathrm{C}$~\cite{azami2024comprehensive, moore1990low}. This is because their glass transition temperatures ($T_\mathrm{g}$) are typically well below these levels, leading to softening and deformation. For example, polylactic acid (PLA) has a $T_\mathrm{g}$ of approximately $55$--$65\,^\circ\mathrm{C}$ and a melting temperature ($T_\mathrm{m}$) around $160$--$170\,^\circ\mathrm{C}$ or slightly above~\cite{marek2016photochemical, srithep2013effects}, while the polyethylene (PE) family exhibits an even lower glass transition temperature ($T_\mathrm{g}$), typically ranging from approximately $-125\,^\circ\mathrm{C}$ to $-73\,^\circ\mathrm{C}$, and a melting point generally between $105$ and $130\,^\circ\mathrm{C}$, depending on its density~\cite{badr2000mechanism, abbott2017thermoplastic, stehling1970glass}.
Under lunar daytime conditions, these materials would easily deform or fail to maintain structural integrity. Another critical limitation is that not all thermoplastics comply with NASA’s strict outgassing requirements for use in vacuum environments. These challenges make high-performance thermoplastics like Polyether–ether–ketone (PEEK) especially attractive for lunar applications. PEEK offers excellent thermal stability ($T_\mathrm{g} = 143\,^\circ\mathrm{C}$, $T_\mathrm{m} = 343\,^\circ\mathrm{C}$), meets NASA’s outgassing standards, and provides additional benefits such as radiation resistance, chemical stability, and abrasion resistance against lunar regolith dust \cite{zanjanijam2020fused, walter1997outgassing, chiggiato2006outgassing, william1993outgassing}. With its high strength-to-weight ratio and proven compatibility with MEX processes, PEEK stands out as a promising material for reliable and efficient lunar manufacturing.

Processing and printing PEEK–regolith composites, however, present several significant challenges. Printing PEEK-based materials is demanding, as PEEK is highly sensitive to temperature fluctuations during fabrication. Variations in thermal conditions can induce thermal stresses, leading to deformation, warpage, and compromised part quality. In addition, PEEK exhibits relatively high melt viscosity, which complicates both the material mixing process and AM process \cite{zanjanijam2020fused, pulipaka2023effect}.

A higher proportion of PEEK in the composite increases material delivery costs, as polymers cannot be synthesized or sourced in situ on the Moon due to the extremely low carbon content in lunar regolith. Therefore, PEEK remains an off-site, Earth-supplied material. Conversely, increasing the weight fraction of regolith in the composite is desirable to reduce reliance on Earth-sourced materials; however, high regolith filler content further increases the melt viscosity of the composite, making extrusion and processing more difficult~\cite{chuang2015additive}. These trade-offs must be carefully considered when designing the printing process and material formulation.

PEEK is well-regarded for its proven performance in space environments and its relatively high potential for recyclability, which can help offset the substantial costs associated with transporting materials from Earth. Although it can withstand more thermal cycles than many other polymers, repeated melting and reprocessing can still degrade its mechanical properties over time \cite{mclauchlin2014studies, pascual2019stability}. Therefore, fabrication strategies should aim to minimize the number of thermal cycles to preserve material integrity and ensure long-term performance.

Efforts have been made to explore the use of MEX AM methods for fabricating regolith–thermoplastic composites. Gelino et al.~\cite{gelino2024selection} investigated several regolith/PLA composite formulations, including weight ratios of 70:30, 80:20, and 85:15 for Lunar Mare simulant Black Point-1 (BP-1), as well as an 80:20 blend of Lunar Highlands Simulant-1 (LHS-1) with PLA. An additional BP-1/PLA mixture incorporating a flow enhancer was also evaluated. These composites were printed under simulated lunar thermal vacuum conditions ($-190~^\circ\mathrm{C}$, $10^{-3}$ torr) and characterized in terms of mixture uniformity, mechanical strength, outgassing behavior, porosity, and density. The 80:20 LHS-1/PLA composite, selected for its favorable mechanical performance and its relevance to the lunar south polar regions, exhibited a flexural modulus of 5.3 GPa and a flexural strength of 24 MPa~\cite{gelino2024selection}. Despite these promising results, as previously discussed, PLA remains inherently unsuitable for uncontained lunar applications because of its poor outgassing behavior and inadequate thermal resistance under lunar environmental conditions.

Our prior work \cite{azami2024additive, azami2025enhancing} reported the first demonstration of manufacturing space-grade high-performance thermoplastic composites with lunar regolith via MEX. That initial work focused on the FFF of PEEK/regolith composites, with comparisons drawn against pure PEEK and PEEK/carbon fiber blends. The study identified that increasing the fraction of solid particles led to extrusion challenges, particularly an increase in sample porosity. The addition of 20~wt\% carbon fiber resulted in an 8.37\% improvement in tensile strength, while incorporating 15~wt\% and 30~wt\% lunar regolith as filler decreased tensile strength by 14.63\% and 26.78\%, respectively. Furthermore, higher regolith content increased brittleness and reduced elongation at break. Microstructural analysis revealed a random dispersion of regolith particles throughout the PEEK matrix, confirming the effectiveness of the mixing approach, but also indicated the presence of pores in both filaments and printed samples. Interestingly, the inclusion of regolith improved interlayer bonding~\cite{azami2024additive, azami2025enhancing}. However, this study served primarily as a proof-of-concept, constrained by the inability at that stage to achieve regolith loadings above 30~wt\% and by the limited scope of characterization.

To address the aforementioned challenges, this study additively manufactured a series of PEEK/regolith composites ranging from pure PEEK to 10~wt\% and increasing in 10~wt\% increments up to 50~wt\% regolith content, using the MEX technique. Pulverized PEEK and lunar regolith were blended and processed into filament form via a twin-screw extruder with a customized screw configuration. The resulting filaments were printed using an FFF MEX system equipped with a heated chamber and a fine-tuned, previously unreported setup, under conditions that enabled in-situ annealing during the printing process. To assess the effects of thermal post-processing, a subset of the printed samples underwent additional annealing, while others remained untreated. The printed composites were evaluated and compared in terms of crystallinity, density, mechanical performance, and microstructural characteristics. To the best of our knowledge, this study reports the highest lunar regolith content incorporated into PEEK to date, using a 50:50~wt\% ratio. PEEK, as a high-performance, high-temperature thermoplastic, was processed via MEX. Specifically, (i) PEEK/regolith composites were developed and successfully printed, (ii) processing parameters were fine-tuned to ensure printability and dimensional fidelity, and (iii) key mechanical, thermal, and microstructural characterizations were conducted. The results highlight both the promise and the practical design limits of regolith-rich thermoplastic composites, representing a significant step toward ISRU for the additive manufacturing of lunar infrastructure, outposts, and functional components.

\section{Materials and Methods}

\subsection{Filament making}

Polyether–ether–ketone (PEEK) pellets (90G, Victrex, United Kingdom) were pulverized using a 254~mm Lab Pulverizer (Orenda, Canada) to achieve an average particle size of approximately $150\,\mu\text{m}$. The resulting PEEK powder was pre-mixed with 10--50~wt\% of as-received lunar regolith simulant (LMS-1D, Exolith, United States), which features a particle size distribution ranging from $<0.04$ to $32\,\mu\text{m}$, with a mean particle size of $7\,\mu\text{m}$, a median (D$_{50}$) of $4\,\mu\text{m}$, D$_{10}$ of approximately $1\,\mu\text{m}$, and D$_{90}$ of $15$--$16\,\mu\text{m}$. According to the supplier, these values were obtained via laser diffraction analysis using a CILAS 1190 particle size analyzer (CILAS, France)~\cite{exolithlms1d}. The simulant has a reported grain density of $2.92\,\text{g/cm}^3$. LMS-1D was selected for its close resemblance to lunar mare dust in terms of particle size distribution, which is a critical parameter influencing composite printability and mechanical performance. The fine particle size aids in achieving uniform dispersion of the regolith within the polymer matrix, thereby reducing stress concentrations and minimizing mechanical failure risks~\cite{nakamura1992effect}.

\begin{figure}[htbp]
    \centering
    \includegraphics[width=1\textwidth]{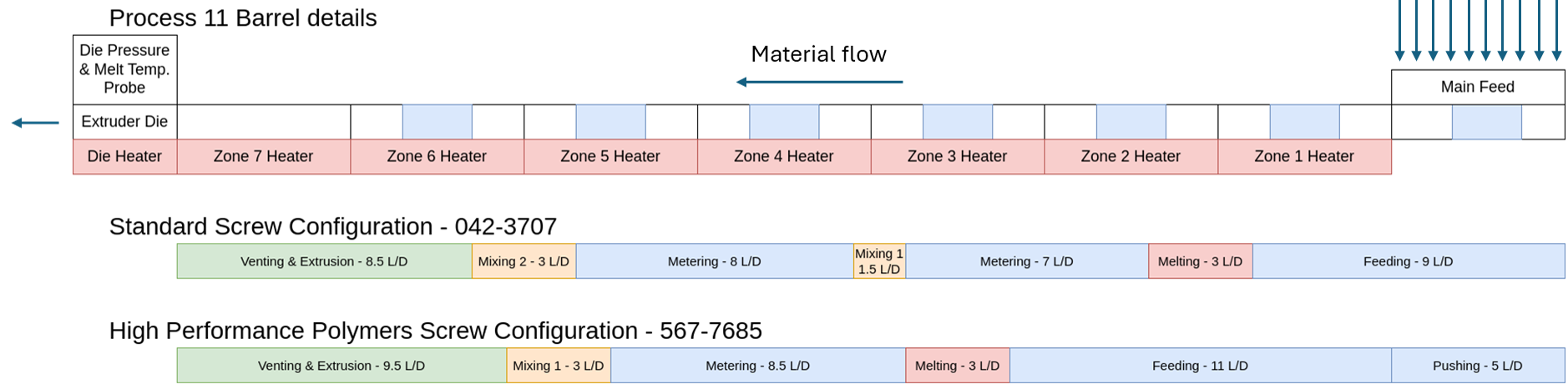}
    \caption{Schematic of the filament making process using the Process 11 twin-screw extruder. The standard screw configuration used in \cite{azami2024additive, azami2025enhancing} is compared with the high performance polymers screw configuration employed in the current study. The updated setup enabled processing of up to 50 wt\% lunar regolith simulant (LRS). L/D denotes the length-to-diameter ratio.}
    \label{screwconfig}
\end{figure}

Compared to earlier work~\cite{azami2024additive, azami2025enhancing}, substantial improvements were made to the filament production process to ensure consistent filament diameter and to increase the extruder's capacity for higher regolith loading. These enhancements included replacing the previously used custom-built filament tensioner with a commercial model, which enabled more stable filament formation and eliminated fluctuations in filament thickness during extrusion. Additionally, a high-performance polymer screw configuration (see Figure \ref{screwconfig}) was implemented in the extrusion system to further improve processing quality. While the prior setup limited regolith content to a maximum of 30~wt\%, the optimized screw design allowed successful processing of filaments containing up to 50~wt\% regolith. 

The PEEK/lunar regolith simulant (PEEK/LRS) composite powder was extruded into filament using a parallel twin-screw extruder (Process~11, Thermo Scientific\texttrademark, Germany), with temperature zones ranging from $355\,^\circ\text{C}$ to $380\,^\circ\text{C}$ and a screw speed of 50--100\,rpm.

For comparative purposes, identical procedures were applied to a commercial PEEK filament (K10, Kexcelled, China), which served as the baseline material in this study.

Prior to printing, all filaments were dried at $65\,^{\circ}\text{C}$ for 24 hours using a filament dryer to minimize moisture content and ensure consistent extrusion quality.

\subsection{Additive manufacturing} 

A high-temperature FFF MEX 3D printer (PEEK~300, CreatBot, China) was utilized to process the prepared composite filaments. To enhance crystallinity and reduce thermal-induced stress, the printer's Direct Annealing System (DAS) was employed, which delivers a controlled stream of hot air at $420\,^\circ\text{C}$ directly to the printed component. Additionally, the build chamber temperature was maintained at $120\,^\circ\text{C}$ to minimize thermal gradients and residual stresses throughout the part. Printing parameters were fine-tuned for each composition to ensure reliable printability and defect-free structures.

In previous work~\cite{azami2024additive, azami2025enhancing}, the following samples were fabricated using the printing parameters summarized in Table~\ref{tab:printing-parameters1}: neat PEEK (\textbf{PEEK}), PEEK reinforced with 20~wt\% chopped carbon fiber (\textbf{PEEK/CF20}), and PEEK composites containing 15~wt\% and 30~wt\% lunar regolith simulant (\textbf{PEEK/LRS15} and \textbf{PEEK/LRS30}, respectively). While acceptable results were achieved, further enhancements in print quality were identified as possible. Consequently, a revised set of printing parameters was applied to a second series of samples, which included neat PEEK and PEEK composites containing 10--50~wt\% LRS. These samples are referred to as \textbf{PEEK}, \textbf{PEEK/LRS10}, \textbf{PEEK/LRS20}, \textbf{PEEK/LRS30}, \textbf{PEEK/LRS40}, and \textbf{PEEK/LRS50} throughout this study. The updated parameters used for their fabrication are summarized in Table~\ref{tab:printing-parameters2}.

For the prior batch of samples~\cite{azami2024additive, azami2025enhancing}, printing was performed on a raft made from the same material as the part, using a single-nozzle configuration. To improve both bed adhesion and bonding between the raft and the printed PEEK/LRS part, a dual-nozzle approach was adopted for this study. In this configuration, the raft, which serves as an intermediate layer between the build plate and the printed component, was fabricated using polyether–ketone–ketone (PEKK). Due to its superior adhesion to the build surface and strong interfacial bonding with PEEK, PEKK significantly reduced detachment during printing and minimized warpage. This improvement enhanced dimensional accuracy and improved the overall structural integrity of the printed parts. The use of PEKK as a dedicated raft material effectively mitigated common challenges in PEEK printing, including delamination, warping, and dimensional inconsistency.

To evaluate reproducibility, three specimens of each composition were printed. In~\cite{azami2024additive, azami2025enhancing}, the surface smoothness of the first set of samples was improved using the ironing feature in UltiMaker Cura; ironing may also reduce residual stress through localized reheating. However, ironing was omitted for the samples in the current study, as it introduced inconsistencies in properties and resulted in nozzle clogging at higher regolith loadings.

%applied at a speed of 5~mm/s and a flow rate of 10\%. This technique involves slow nozzle traversal at elevated temperatures, which smooths the outer surface and 

\begin{table}[ht]
\caption{Summary of the printing parameters used in Azami et al.~\cite{azami2024additive, azami2025enhancing}.}
\label{tab:printing-parameters1}
\centering
\footnotesize
\setlength{\tabcolsep}{3pt}
\small
\begin{tabular}{l c c c c c c c c}
\hline
\multicolumn{1}{l}{\begin{tabular}[c]{@{}c@{}}Material\end{tabular}} &
\multicolumn{1}{c}{\begin{tabular}[c]{@{}c@{}}Nozzle\\ Temp.\\ (°C)\end{tabular}} &
\multicolumn{1}{c}{\begin{tabular}[c]{@{}c@{}}Bed\\ Temp.\\ (°C)\end{tabular}} &
\multicolumn{1}{c}{\begin{tabular}[c]{@{}c@{}}Ambient\\ Temp.\\ (°C)\end{tabular}} &
\multicolumn{1}{c}{\begin{tabular}[c]{@{}c@{}}Layer\\ Thickness\\ (mm)\end{tabular}} &
\multicolumn{1}{c}{\begin{tabular}[c]{@{}c@{}}Print\\ Speed\\ (mm/s)\end{tabular}} &
\multicolumn{1}{c}{\begin{tabular}[c]{@{}c@{}}Infill \\Density\\ (\%)\end{tabular}} &
\multicolumn{1}{c}{\begin{tabular}[c]{@{}c@{}}Nozzle \\Diameter\\ (mm)\end{tabular}} &
\multicolumn{1}{c}{\begin{tabular}[c]{@{}c@{}}Bed \\Adhesion\\ Mechanism\end{tabular}} \\
\hline
PEEK & 420 & 180 & 120 & 0.25 & 7.5 & 100 & 0.4 & PEEK Raft \\
PEEK/CF20 & 400 & 150 & 120 & 0.25 & 20 & 100 & 0.4 & None \\
PEEK/LRS15 & 420 & 180 & 120 & 0.25 & 7.5 & 100 & 0.4 & PEEK/LRS15 Raft \\
PEEK/LRS30 & 420 & 180 & 120 & 0.25 & 7.5 & 100 & 0.4 & PEEK/LRS30 Raft \\
\hline
\end{tabular}
\normalsize
\end{table}

\begin{table}[ht]
\caption{Summary of the printing parameters used in the current study.}
\label{tab:printing-parameters2}
\centering
\footnotesize
\setlength{\tabcolsep}{3pt}
\small
\begin{tabular}{l c c c c c c c c}
\hline
\multicolumn{1}{l}{\begin{tabular}[c]{@{}c@{}}Material\end{tabular}} &
\multicolumn{1}{c}{\begin{tabular}[c]{@{}c@{}}Nozzle\\ Temp.\\ (°C)\end{tabular}} &
\multicolumn{1}{c}{\begin{tabular}[c]{@{}c@{}}Bed\\ Temp.\\ (°C)\end{tabular}} &
\multicolumn{1}{c}{\begin{tabular}[c]{@{}c@{}}Ambient\\ Temp.\\ (°C)\end{tabular}} &
\multicolumn{1}{c}{\begin{tabular}[c]{@{}c@{}}Layer\\ Thickness\\ (mm)\end{tabular}} &
\multicolumn{1}{c}{\begin{tabular}[c]{@{}c@{}}Print\\ Speed\\ (mm/s)\end{tabular}} &
\multicolumn{1}{c}{\begin{tabular}[c]{@{}c@{}}Infill \\Density\\ (\%)\end{tabular}} &
\multicolumn{1}{c}{\begin{tabular}[c]{@{}c@{}}Nozzle \\Diameter\\ (mm)\end{tabular}} &
\multicolumn{1}{c}{\begin{tabular}[c]{@{}c@{}}Bed \\Adhesion\\ Mechanism\end{tabular}} \\
\hline
PEEK & 430 & 180 & 120 & 0.25 & 7.5 & 100 & 0.4 & PEEK Raft \\
PEEK/LRS & 430 & 180 & 120 & 0.25 & 7.5 & 100 & 0.4 & PEKK Raft \\
\hline
\end{tabular}
\normalsize
\end{table}

\subsection{Annealing heat treatment}

Although the printer's DAS provided in-situ annealing during the printing process, an additional post-processing annealing step was carried out on a subset of samples to assess its influence on the properties of the as-printed parts. These specimens will be referred to as ``annealed'' throughout the remainder of this article. The annealing procedure was conducted in a furnace, where the sample temperature was increased at a controlled rate of 5~$^\circ$C/min until reaching 300~$^\circ$C. The samples were held at this temperature for 2~hours, followed by gradual cooling inside the furnace by simply turning off the heating element and allowing the system to cool down naturally.

\subsection{Differential scanning calorimetry (DSC)}

Thermal behavior was analyzed using a TA Instruments DSC Q200 equipped with \textit{Universal Analysis} V4.7A software (TA Instruments, USA). Each specimen, cut from samples, was sealed in a standard aluminum pan and run against an empty reference under a dry nitrogen purge (50~mL~min$^{-1}$). The thermal program consisted of a single heating and cooling cycle: samples were stabilized at $50\,^\circ\mathrm{C}$, heated to $450\,^\circ\mathrm{C}$ at $10\,^\circ\mathrm{C}\,\mathrm{min}^{-1}$, and immediately cooled back to $50\,^\circ\mathrm{C}$ at the same rate. No second heating was performed. Various compositions of materials were examined, ranging from neat PEEK to composites containing up to 50~wt\% regolith. 

%The objective of the test was to investigate how the presence and content of regolith influence the PEEK crystallization of the as-printed parts, and to identify the temperature interval that yields the highest crystallinity, potentially useful for post-printing thermal treatment.

The degree of crystallinity of the polymer phase ($\chi$) was calculated based on the net enthalpy of fusion, following the approach adopted by Yap et al.~\cite{yap2023additive}:

\begin{equation}
  \chi = 100\% \times
         \left(
           \frac{\Delta H_m + \Delta H_c}
                {\Delta H_f \,(1 - w_f)}
         \right),
  \label{eq:crystallinity}
\end{equation}

In this equation, $\Delta H_m$ refers to the enthalpy of fusion obtained from the melting endotherm (J\,g$^{-1}$), while $\Delta H_c$ is the magnitude of the crystallization exotherm (J\,g$^{-1}$). The reference value $\Delta H_f = 130$\,J\,g$^{-1}$ corresponds to the enthalpy of fusion for fully crystalline PEEK~\cite{blundell1983morphology}, and $w_f$ represents the mass fraction of regolith filler. All enthalpy values were initially normalized to the total mass of the composite; division by $(1 - w_f)$ subsequently converts them to a polymer-phase basis, thereby correcting for the dilution effect introduced by the inert filler.

\subsection{Tensile examination} 

For tensile testing, all specimens were prepared in accordance with the ISO~527 standard. The tests were performed using a 5000~N universal testing machine (Hoskin Scientific, Canada) at a constant crosshead speed of 2~mm/min.

\subsection{Density assessment} 

The density of both the filaments and the printed samples was measured using Archimedean density analysis, employing a density determination kit (YDK02MS, Sartorius, Germany) with deionized water at $22\,^\circ\text{C}$ as the immersion medium.

\subsection{Micrography} 

Microstructural characterization of both the filaments and as-printed specimens was performed using a scanning electron microscope (SEM, S-3400N, Hitachi, Japan). Prior to imaging, cross-sectional samples were polished to enable high-resolution surface examination. Elemental composition was analyzed using energy-dispersive X-ray spectroscopy (EDS). To ensure a comprehensive assessment, EDS mapping was also conducted over broad regions rather than isolated points. Continuous signal acquisition across the selected areas allowed the generation of spatially resolved elemental distribution maps, providing insights into the homogeneity and dispersion of the composite constituents.

\section{Results and Discussion}

\subsection{Filament preparation}

\begin{figure}[htbp]
    \centering
    \includegraphics[width=0.5\textwidth]{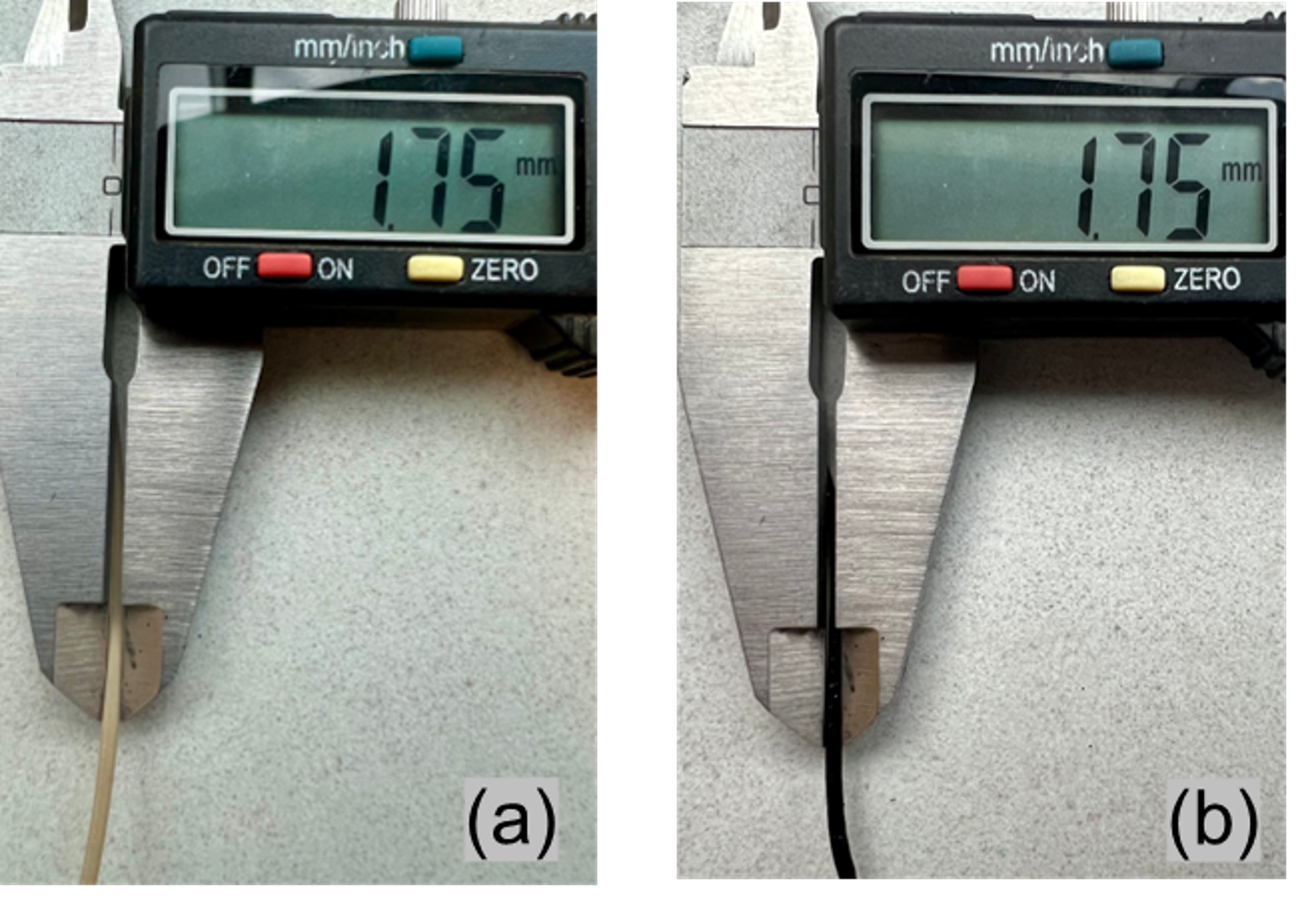}
    \caption{Filament comparison: a) Neat PEEK, b) PEEK/LRS30}
    \label{filaments}
\end{figure}

As detailed in \cite{azami2024additive, azami2025enhancing}, the use of pulverized PEEK, combined with an increased screw speed of 100~rpm and a maintained processing temperature of $380\,^\circ\mathrm{C}$, enabled successful processing of the PEEK/regolith composite using a twin-screw extruder. This approach yielded a uniform filament (other than some short-lived inconsistencies), with the extruder’s high shear mixing capability playing a crucial role in achieving homogeneous dispersion of the regolith particles. This was further supported by the stable torque observed throughout the extrusion process. Figure~\ref{filaments} presents an example comparison between the neat PEEK filament and the composite filament containing 30~wt\% regolith.

\begin{figure}[htbp]
    \centering
    \includegraphics[width=0.4\textwidth]{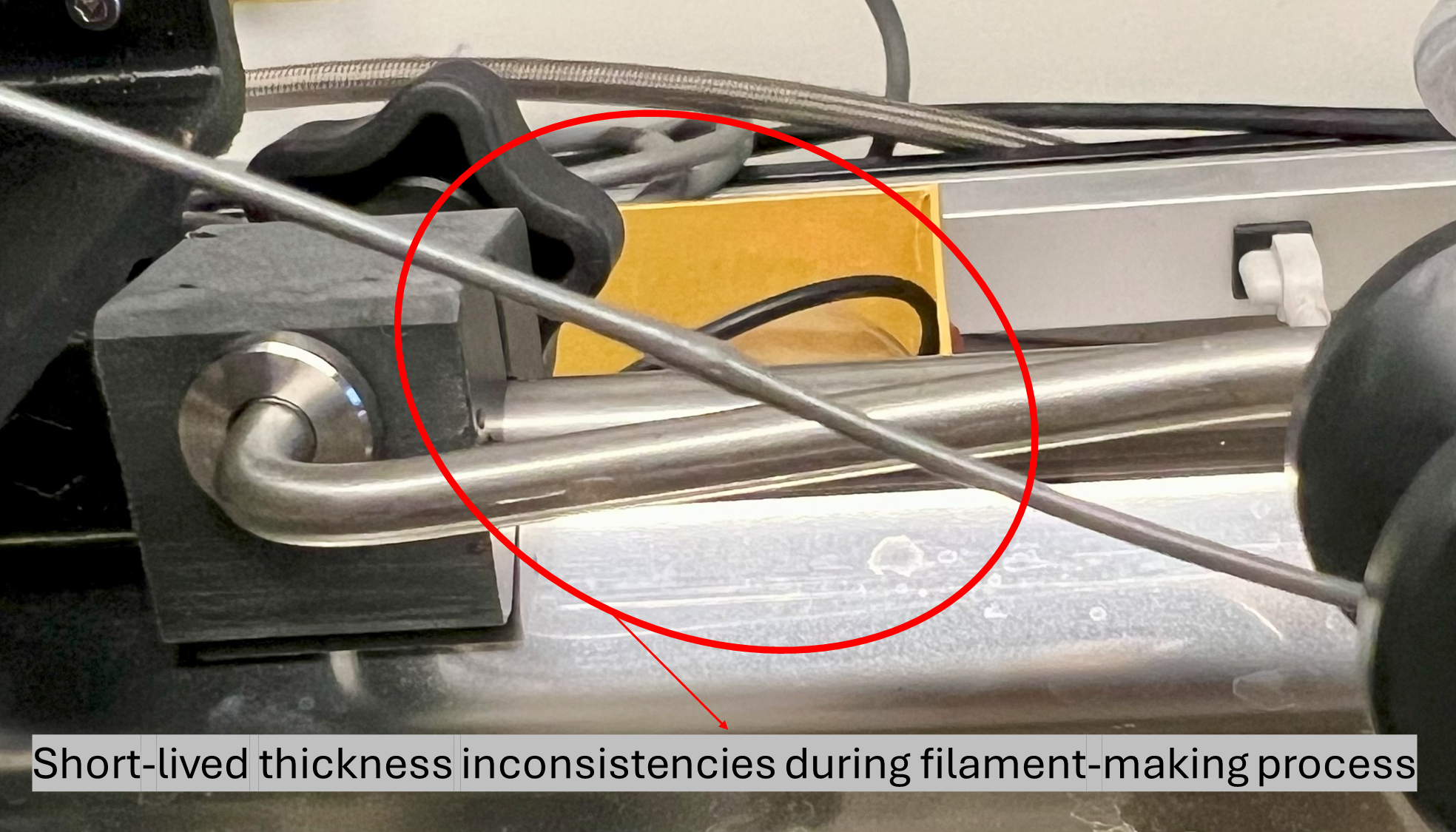}
    \caption{Example of intermittent filament thickness inconsistency ranging from 1 to 4 cm during extrusion before modification.}
    \label{fig:3}
\end{figure}

However, increasing the regolith content beyond 30~wt\% resulted in a significant rise in torque, which adversely affected the extrusion process. A 10--15\% increase in torque was already observed when the regolith content increased from 15~wt\% to 30~wt\%, primarily due to the elevated melt viscosity of the mixture. This viscosity increase is attributed to the higher proportion of solid particles that remain unmelted during processing, as reported in~\cite{chuang2015additive}. At contents exceeding 30~wt\%, the torque rose further, making continuous extrusion increasingly difficult. Additionally, intermittent inconsistencies in filament diameter, typically spanning 1--4\,cm in length, were occasionally observed. These defects required manual intervention to remove the affected segments and resume spooling. An example of such an inconsistency is shown in Figure~\ref{fig:3}.

Subsequent improvements implemented in this work included replacing the custom-built filament tensioner with a commercial unit and adopting the ``high-performance polymer'' screw configuration recommended by the twin-screw extruder’s manufacturer (Thermo Scientific\texttrademark, Germany). These modifications collectively resulted in a 30--40\% reduction in torque during extrusion compared to the original configuration. Notably, this setup enabled the successful extrusion of composite blends containing up to 50~wt\% regolith. This was achieved through a combination of the improved screw configuration, a 30\% reduction in screw speed, and a $10\,^\circ\mathrm{C}$ increase in processing temperature across all extruder zones. Remarkably, the torque recorded during extrusion of the 50~wt\% blend was over 10\% lower than that observed for the 30~wt\% blend processed with the initial configuration. Although processing even higher regolith contents may be technically feasible, the blend ratio was intentionally limited to 50~wt\% as a precaution against any potential accumulative negative effects on the mechanical integrity of the extruder.

\subsection{Additive manufacturing}

Following filament preparation, a range of printing parameters, including nozzle temperature, bed temperature, and printing speed, were systematically fine-tuned to enhance the quality of the printed specimens. Any parts exhibiting visible defects such as cracking or excessive porosity were excluded from further analysis. The specimens fabricated using the fine-tuned parameters listed in Table~\ref{tab:printing-parameters2} displayed no observable defects.

Print consistency and process repeatability were confirmed through visual inspection, tensile testing, density measurements, and dimensional analysis. These evaluations indicate that the refined parameters produced reliable and reproducible results across different material compositions.

A key finding of this study is that incorporating regolith into the polymer matrix significantly reduced delamination and warping, which are common challenges in pure PEEK prints, resulting in enhanced dimensional accuracy and higher print success rates.

\begin{figure}[htp]
    \centering
    \includegraphics[width=6cm]{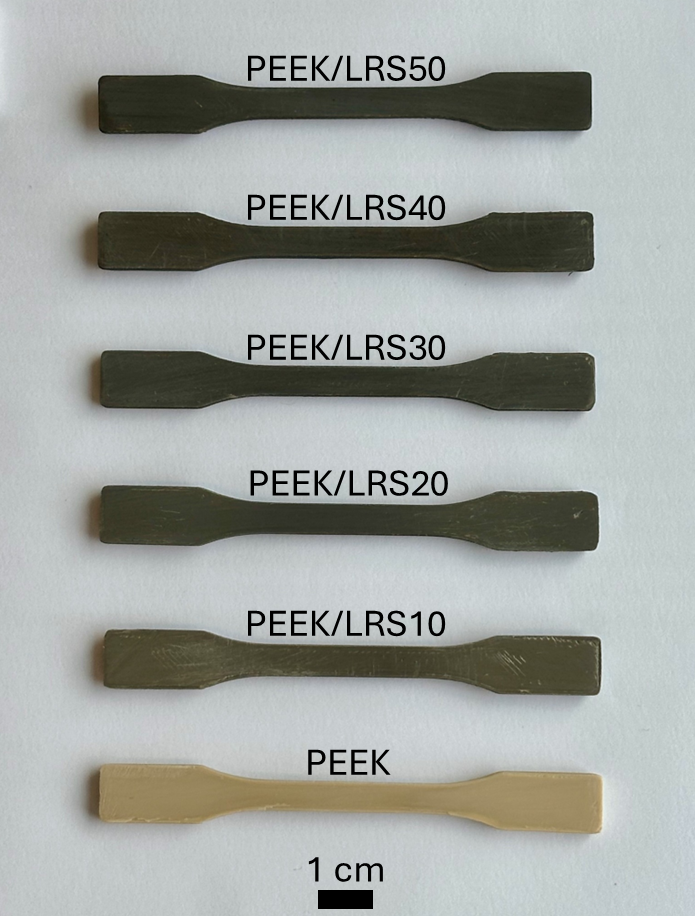}
    \caption{Representative examples of as-printed replicas, ranging from neat PEEK to PEEK with 50~wt\% regolith.}
    \label{sampledcom}
\end{figure}

To further enhance print quality, PEKK was introduced as the raft material in place of PEEK/LRS composites. This modification proved critical for print reliability. When PEEK-based rafts (composed of the same material as the printed part) were used, success rate was less than 50\%. In contrast, substituting PEKK as the raft material resulted in an almost 100\% success rate. This outcome underscores the importance of interfacial adhesion, both between the raft and the print bed and between the raft and the printed structure. Acting as an effective intermediary, PEKK provided superior bonding at both interfaces, thereby eliminating adhesion-related failures. This improvement represents a significant advancement in the additive manufacturing of PEEK-regolith composites, especially in the context of space-based fabrication, where minimizing material waste and maximizing print reliability is critical. Representative samples from this phase are shown in Figure~\ref{sampledcom}.

\subsection{Effect of annealing}

Upon annealing, no notable changes were observed in the color or overall dimensions of the samples, aside from a slight and uniform shrinkage likely associated with pore consolidation. The parts retained their structural integrity. The effects of annealing on other material properties are discussed in the corresponding sections.

\subsection{Density analysis}

The relative density of the feedstock filaments, as-printed specimens, and annealed samples for neat PEEK as well as all PEEK/LRS composites was determined using the Archimedean method and reported as a percentage of the corresponding nominal theoretical density (see Table~\ref{density}). It was observed that, for all compositions and manufacturing strategies, the variation in relative density across replicates remained below 3\%. To facilitate a clearer interpretation of compositional trends, the values reported here correspond to the samples exhibiting median tensile strength for each composition.

For pure \textbf{PEEK}, the filament exhibited a high density of 99.0\%, while both the as-printed and annealed specimens maintained densities exceeding 99\%. These results confirm that the selected thermal and processing parameters, particularly the nozzle temperature and chamber conditioning, were effective in promoting sufficient melt flow and inter-raster bonding during printing, yielding a slight improvement over the outcomes reported in~\cite{azami2025enhancing}.

For the \textbf{PEEK/LRS10} composite, the filament exhibited a relative density of 97.8\%. Printing improved this to 98.5\%, indicating that the melt pressure and thermal conditions were sufficient to overcome the moderate viscosity increase due to filler addition. Annealing at $300\,^\circ\mathrm{C}$ further increased the density to above 99\%, highlighting the effectiveness of post-processing in eliminating residual porosity.

The \textbf{PEEK/LRS20} filament showed a slightly lower density of 97.0\%, attributed to increased melt viscosity with higher filler content. After printing, the density increased to 97.6\%, and subsequent annealing restored it to above 99\%, confirming that the matrix retained sufficient mobility for defect healing and densification.

For the \textbf{PEEK/LRS30} composite, filament density dropped further to 96.1\%, and printing offered only a marginal improvement (96.2\%), indicating that inter-raster fusion was hindered by the higher viscosity. Annealing yielded a modest increase to 97.2\%, reflecting the diminishing effectiveness of thermal post-treatment at higher filler contents.

The \textbf{PEEK/LRS40} filament exhibited a higher density (98.2\%) than PEEK/LRS30, which is attributed to process modifications during extrusion: the twin-screw rpm was reduced by half, allowing more time for air evacuation, and the increased solid content elevated die pressure, enhancing melt compaction. Nevertheless, the as-printed density dropped to 95.1\% (–3.1\%), primarily due to inadequate flow under the high-viscosity conditions. Annealing offered limited improvement, increasing density by only 0.5\%.

In the \textbf{PEEK/LRS50} case, filament density declined to 96.9\% (–1.3\% compared to PEEK/LRS40). High viscosity prevented adequate melt pressure and flow continuity during printing, leading to a substantial reduction in as-printed density to 92.5\% (–4.5\% compared to filament). Annealing resulted in only a 0.3\% gain, underscoring the limited mobility of the polymer chains within the highly constrained matrix.

\textbf{Trend analysis.} Up to 30~wt\% regolith, filament porosity increases progressively with filler content due to rising melt viscosity. At 40~wt\% there is a discontinuity in the trend, with improved filament density (i.e. a drop in porosity) resulting from longer residence time and higher die pressure during extrusion. At 50~wt\%, porosity increases again, likely due to the limits of degassing efficiency under extreme viscosity. Further reduction in screw speed might have improved air evacuation, but was avoided to prevent material degradation due to prolonged residence in the high-temperature barrel. For regolith contents of 10~wt\% to 30~wt\%, porosity decreases during printing (relative to filament porosity), as the process conditions effectively promote interbead diffusion. Beyond this point in regolith content, the trend reverses: the as-printed porosity increases sharply due to poor flow and inadequate bead fusion at high filler loadings. Further increasing the nozzle temperature was not pursued, as the selected setting ($430\,^\circ\mathrm{C}$) approaches the onset of thermal degradation in PEEK.

Post-processing annealing was effective for composites with 10~wt\% and 20~wt\%, closing interfacial voids. However, as regolith content increases further, the benefits diminish considerably. This trend is attributed to reduced chain mobility: at the annealing temperature of $300\,^\circ\mathrm{C}$ (approximately $44\,^\circ\mathrm{C}$ below the melting point of PEEK), polymer segments must diffuse around rigid filler particles. As the filler content increases, the available free matrix volume decreases, limiting molecular rearrangement and the potential for crystallinity development, weld healing, and porosity reduction. Consequently, the mechanical enhancement achievable through annealing becomes increasingly constrained at higher regolith loadings.

\begin{table}[ht]
\caption{Relative density (\%) of pure PEEK and PEEK--regolith (PEEK/LRS) composites at three processing stages: filament, as-printed, and annealed. Columns 5--6 show percent changes between the stages; Columns 7--9 show percent changes relative to the previous filler loading. Values represent the average of three specimens. \textit{Note:} Extrusion parameters changed between 30 and 40~wt\% LRS.} \label{density}
\label{tab:rel-density-change}
\centering
\begin{adjustbox}{width=\textwidth}
\begin{tabular}{l c c c c c c c c}
\hline
\textbf{Regolith Content (wt\%)} &
\textbf{Filament (\%)} &
\textbf{As-Printed (\%)} &
\textbf{Annealed (\%)} &
\textbf{\%$\Delta$ Fil$\rightarrow$Pr} &
\textbf{\%$\Delta$ Pr$\rightarrow$Ann} &
\textbf{\%$\Delta$ Fil vs Prev} &
\textbf{\%$\Delta$ Pr vs Prev} &
\textbf{\%$\Delta$ Ann vs Prev} \\
\hline
0 (Pure PEEK) & $>99$ & $>99$ & $>99$ & --        & --        & --       & --       & --        \\
10            & 97.8  & 98.5  & $>99$ & +0.72\%   & $\ge$+0.51\% & --       & --       & --        \\
20            & 97.0  & 97.6  & $>99$ & +0.62\%   & $\ge$+1.44\% & --0.82\% & --0.91\% & --        \\
30            & 96.1  & 96.2  & 97.2  & +0.10\%   & +1.04\%      & --0.93\% & --1.43\% & --        \\
40            & 98.2*  & 95.1  & 95.6  & --3.16\%  & +0.53\%      & +2.19\%*  & --1.14\%* & --1.65\%*  \\
50            & 96.9  & 92.5  & 92.8  & --4.54\%  & +0.32\%      & --1.32\% & --2.73\% & --2.93\%  \\
\hline
\end{tabular}
\end{adjustbox}
\normalsize
\end{table}

\iffalse
\begin{table}[ht]
\caption{Average Relative Density (\%) of Pure PEEK and PEEK–Regolith Composites in Different Processing Stages. Values represent the average of three specimens.}
\label{tab:rel-density}
\centering
\footnotesize
\setlength{\tabcolsep}{6pt}
\small
\begin{tabular}{l c c c}
\hline
\multicolumn{1}{l}{\textbf{Regolith Content (wt\%)}} &
\multicolumn{1}{c}{\textbf{Filament}} &
\multicolumn{1}{c}{\textbf{As-Printed}} &
\multicolumn{1}{c}{\textbf{Annealed}} \\
\hline
0 (Pure PEEK)   & $>99$     & $>99$  & $>99$  \\
10              & 97.8      & 98.5   & $>99$  \\
20              & 97.0      & 98.0   & $>99$  \\
30              & 96.1      & 96.2   & 97.2   \\
40              & 98.2      & 95.1   & 95.6   \\
50              & 96.9      & 92.5   & 92.8   \\
\hline
\end{tabular}
\normalsize
\end{table}
\fi

\subsection{Differential scanning calorimetry (DSC)}

\begin{figure}[htbp]
    \centering

    % Row 1: (a) and (b)
    \begin{subfigure}[b]{0.48\textwidth}
        \centering
        \includegraphics[width=\linewidth]{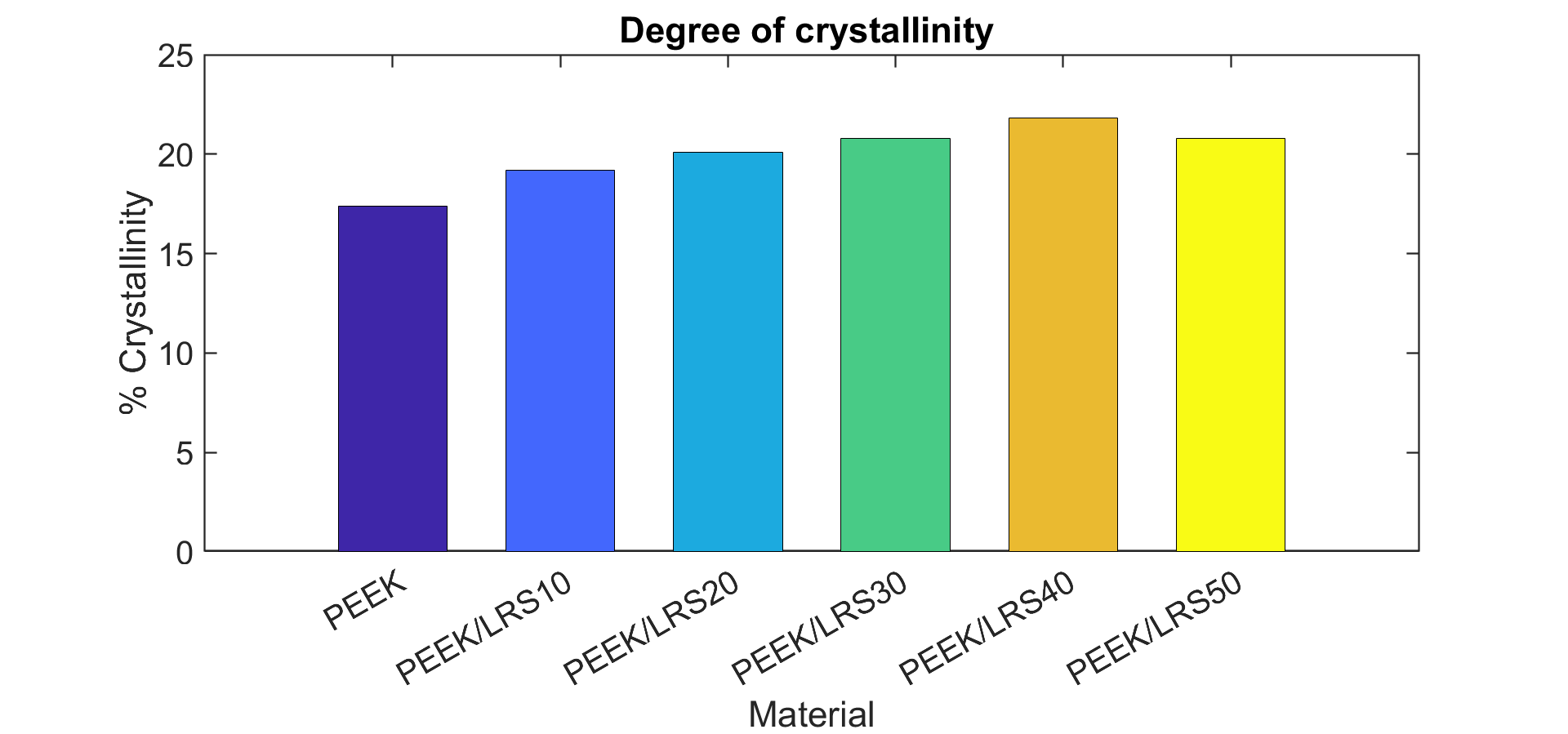}
        \caption{Degree of crystallinity.}
        \label{fig:crystallinity_a}
    \end{subfigure}
    \hfill
    \begin{subfigure}[b]{0.48\textwidth}
        \centering
        \includegraphics[width=\linewidth]{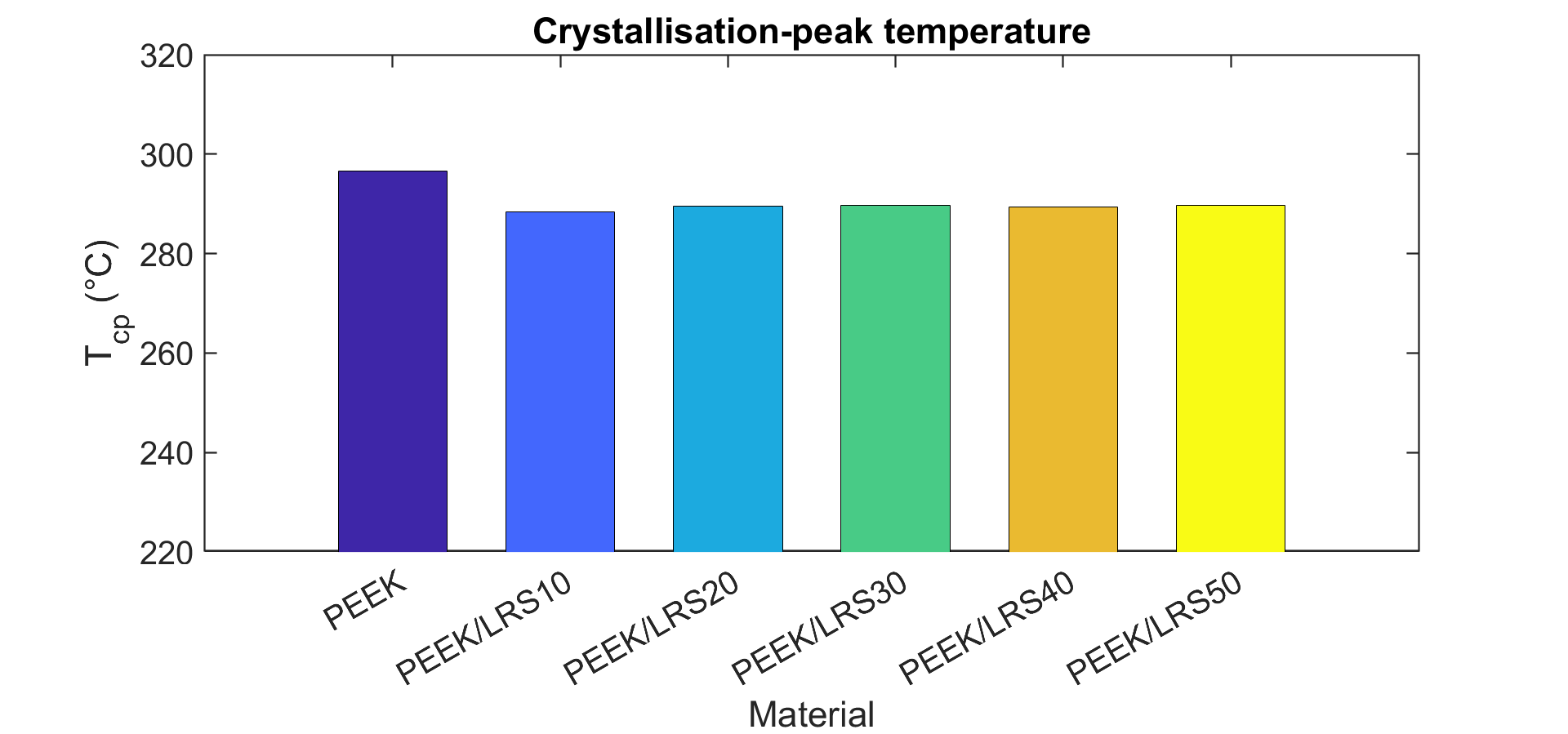}
        \caption{Crystallization-peak temperature, \(T_{\mathrm{cp}}\).}
        \label{fig:crystallinity_b}
    \end{subfigure}

    \vskip\baselineskip

    % Row 2: (c) and (d)
    \begin{subfigure}[b]{0.48\textwidth}
        \centering
        \includegraphics[width=\linewidth]{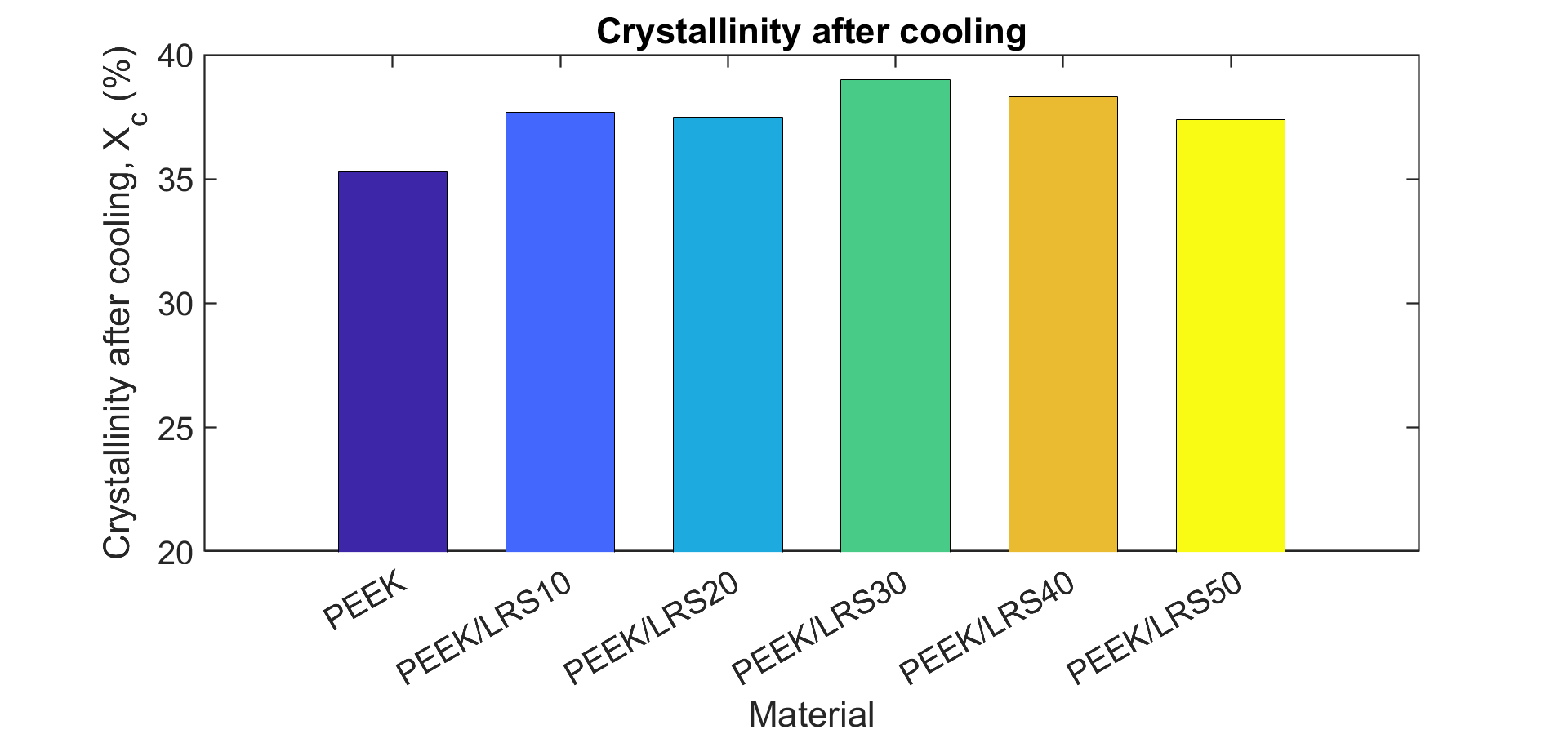}
        \caption{Crystallinity after cooling, \(X_c\).}
        \label{fig:crystallinity_c}
    \end{subfigure}
    \hfill
    \begin{subfigure}[b]{0.48\textwidth}
        \centering
        \includegraphics[width=\linewidth]{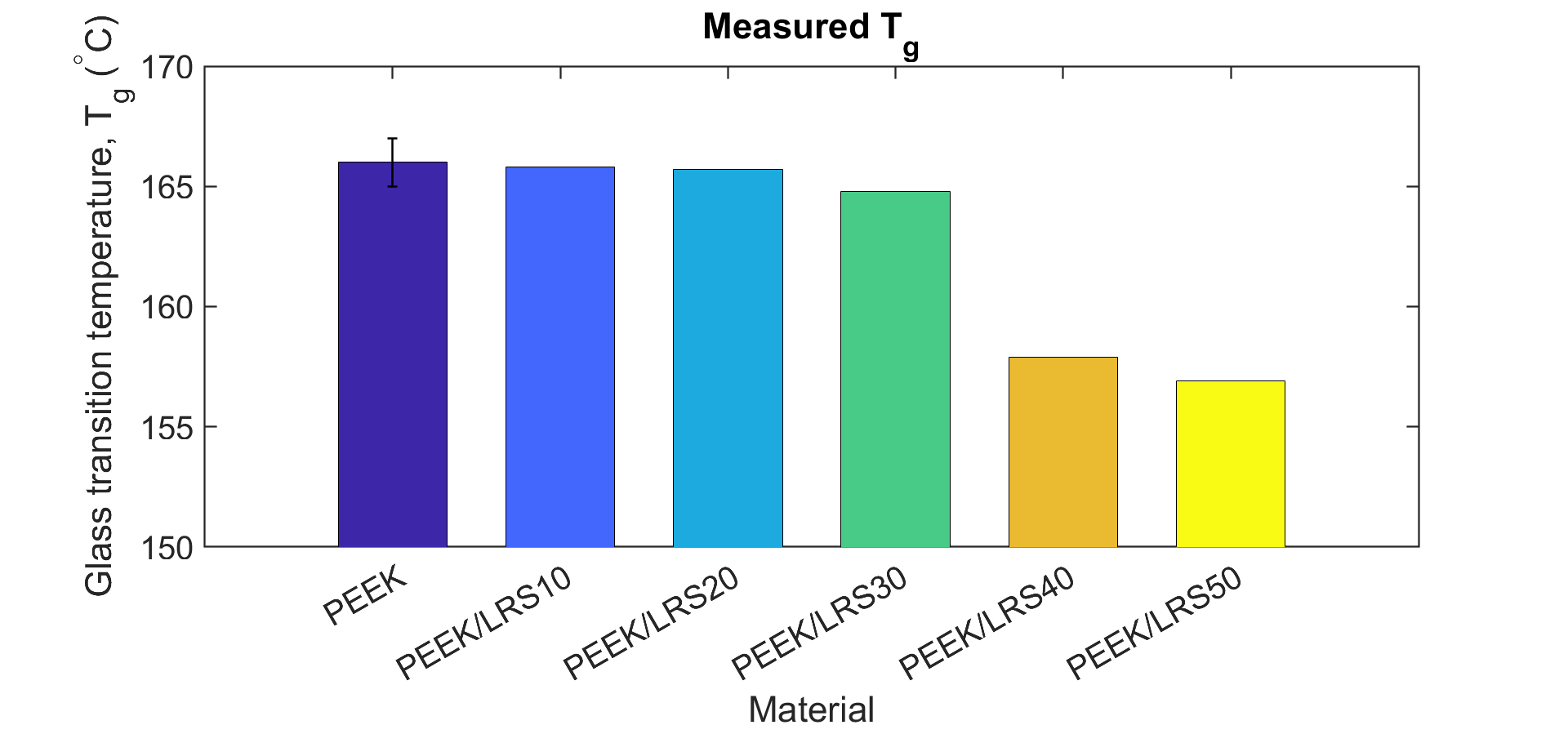}
        \caption{Glass transition temperature, \(T_g\).}
        \label{fig:crystallinity_d}
    \end{subfigure}

    \caption{DSC-derived crystallization and thermal behavior of as-printed PEEK and PEEK/LRS composites.}
    \label{fig:crystallinity_summary}
\end{figure}

As shown in Figure~\ref{fig:crystallinity_a}, DSC revealed that the as-printed neat PEEK exhibited a degree of crystallinity of \(17.4\%\). Incorporation of regolith increased the crystallinity to \(20.5 \pm 1.3\%\) across the composites, with a gradual increase up to 40~wt\%, followed by a slight decrease at 50~wt\%. The regolith particles, particularly those at the nano- and submicron scales, serve as effective heterogeneous nucleation sites, lowering the energy barrier for crystallization during cooling. This promotes rapid nucleation and leads to the formation of numerous fine crystal grains. However, in the presence of filler, the crystallization peak temperature (\(T_{\mathrm{cp}}\)) shifts to lower values, decreasing from 296.6\,\(^\circ\mathrm{C}\) for pure PEEK to \(289.1 \pm 0.7\,^\circ\mathrm{C}\) for the PEEK/LRS composites (see Figure~\ref{fig:crystallinity_b}). This downward shift indicates an increasing restriction in PEEK chain mobility during solidification. In this regime, the reduced molecular mobility becomes the dominant influence, limiting crystal growth and slightly decreasing both \(T_{\mathrm{cp}}\) and the overall crystallinity (\(X_{\mathrm{c}}\)). Consequently, at high filler contents, the inhibitory effect on chain mobility outweighs the nucleation-enhancing benefits of the particles \cite{lai2007peek}.

\begin{figure}[htbp]
    \centering
    \includegraphics[width=0.7\textwidth]{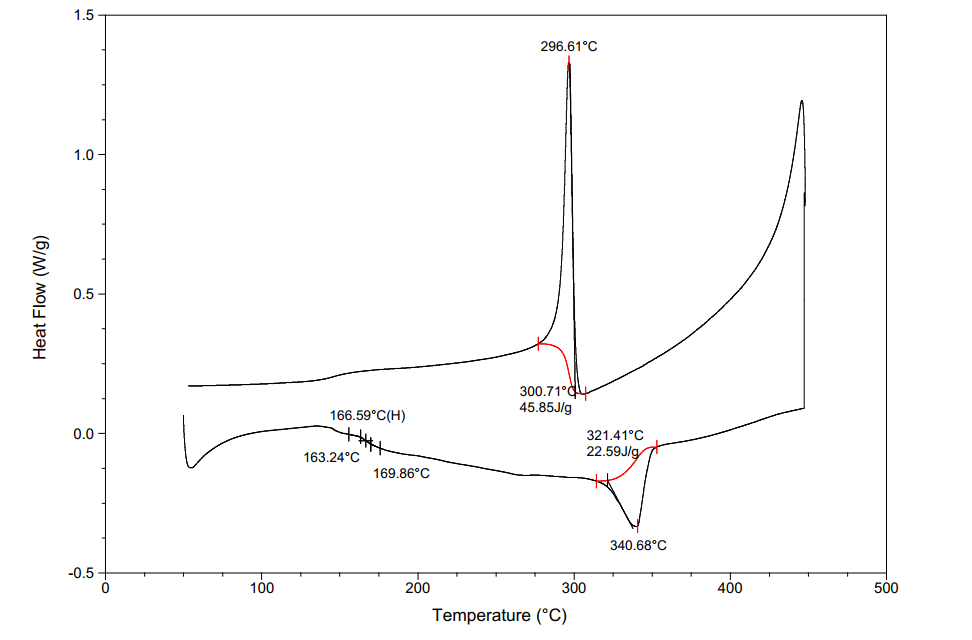}
    \caption{Representative DSC thermogram (heating and cooling cycle) of as-printed neat PEEK.}
    \label{fig:DSC}
\end{figure}

A representative DSC thermogram is provided in Figure \ref{fig:DSC} to illustrate the typical thermal behavior, while key thermal parameters for various composition are summarized in Figure~\ref{fig:crystallinity_summary}.

The crystallization exotherms recorded during the cooling cycle revealed that the neat PEEK sample achieved a crystallinity of 35.3\%, while all PEEK/LRS composites exhibited comparable values, averaging \(38.3 \pm 0.7\%\) (see Figure~\ref{fig:crystallinity_c}). This consistency is primarily attributed to the fixed cooling rate of \(10\,^\circ\mathrm{C\,min^{-1}}\), which ensured uniform and sufficiently slow thermal conditions across all samples. For neat PEEK, the relatively high crystallization peak temperature \((T_{\mathrm{cp}} = 296.6\,^\circ\mathrm{C})\) and high chain mobility provided adequate time for lamellar growth. In contrast, the PEEK/LRS composites underwent rapid heterogeneous nucleation due to the abundance of regolith particles, resulting in an average increase of 3.0 percentage points in crystallinity (from 35.3\% to 38.3\%). At these higher crystallinity levels, the process becomes growth-limited rather than nucleation-limited. Once nucleation sites are saturated, further increases in filler content no longer enhance crystallinity, and the final degree of crystallinity is largely governed by the imposed cooling conditions.

The glass transition temperature ($T_\mathrm{g}$) of as-printed neat PEEK was measured to be \(166 \pm 1\,^\circ\mathrm{C}\), based on four repeated tests. Notably, these values are significantly higher than the typical $T_\mathrm{g}$ reported in data sheets (onset at 143\,$^\circ$C and midpoint at 150\,$^\circ$C), which can be attributed to the elevated thermal conditions applied during printing. Specifically, Creatbot's DAS was employed, wherein both tuyeres directed hot air at 420\,$^\circ$C, complemented by a heated bed at 180\,$^\circ$C, and a build chamber maintained at 120\,$^\circ$C. Similar $T_\mathrm{g}$ elevations have also been reported in previous studies~\cite{arzak1991effect, lannunziata2024effect}. The observed increase in $T_\mathrm{g}$ results from a combination of crystallization and thermally induced changes in the amorphous phase. While annealing is commonly associated with enhancements in crystalline structure and degree of crystallinity in semi-crystalline polymers such as PEEK, it also significantly reduces chain mobility within the rigid amorphous regions. This reduced mobility contributes to the observed elevation in $T_\mathrm{g}$~\cite{jiang2019effect, lannunziata2024effect}.

As shown in Figure~\ref{fig:crystallinity_d}, the glass transition temperature (\(T_\mathrm{g}\)) remained stable at \(165.2 \pm 0.5\,^\circ\mathrm{C}\) for composites containing up to 30~wt\% LRS. However, it decreased noticeably to \(157.9\,^\circ\mathrm{C}\) and \(156.9\,^\circ\mathrm{C}\) for the 40~wt\% and 50~wt\% LRS composites, respectively. This reduction is attributed to the increased incidence of processing-induced defects and increased porosity in as-printed parts at higher filler loadings, which elevate the free volume and enhance segmental mobility within the polymer matrix~\cite{Lau2025_TA}.

For the annealed samples, a modest increase in crystallinity was observed for neat PEEK, rising from 17.4\% to 20.1\%. In contrast, no meaningful change was detected in the DSC data of the PEEK/LRS composites compared to the as-printed specimens. DSC analysis indicated that annealing at \(300\,^\circ\mathrm{C}\) should be sufficient to promote crystallization in PEEK without risking part deformation. Specifically, the cooling curves of the as-printed samples display a pronounced exothermic crystallization peak ending just below \(300\,^\circ\mathrm{C}\), confirming that this temperature lies within the effective crystallization window. Meanwhile, the heating curves reveal that melting onset begins around \(321\,^\circ\mathrm{C}\), suggesting that annealing above \(300\,^\circ\mathrm{C}\) would approach the deformation threshold. Therefore, holding the parts at \(300\,^\circ\mathrm{C}\) should, in principle, enable further crystal growth while maintaining dimensional stability, making it a practical and thermally safe annealing temperature.

This annealing condition is more aggressive than many protocols reported in the literature. For example, Lannunziata et al.~\cite{lannunziata2024effect} and Sarasua et al.~\cite{sarasua1996effects} used temperatures as low as \(200\,^\circ\mathrm{C}\). However, no appreciable increase in the crystallinity of the PEEK/LRS samples was observed. One contributing factor may be the partial crystallization that occurred during the printing process itself. The slow-cooling environment, characterized by a high bed and ambient temperature, and the simultaneous annealing enabled by the DAS system facilitated primary crystal formation during printing. Once a lamellar network is established, further crystal growth requires long-range chain diffusion, which becomes increasingly limited under these conditions.

Additionally, in the PEEK/LRS composites, the regolith particles disrupt the continuity of the polymer matrix by creating isolated domains and interfacial layers where chain mobility is suppressed. These particles act as barriers to lamellar propagation, restricting crystallization to already available polymer-rich regions. Furthermore, the presence of interlayer gaps in MEX-printed PEEK introduces microvoids and poorly bonded interfaces, which reduce chain continuity and hinder the nucleation and growth of crystalline domains.

Although slightly higher annealing temperatures (e.g., \(315{-}320\,^\circ\mathrm{C}\)) might enhance recrystallization, they also pose a higher risk of deformation. In the context of space applications, such thermal treatments may be impractical due to additional energy, equipment, and time demands. Thus, while \(300\,^\circ\mathrm{C}\) is thermodynamically favorable for crystallization, the combination of kinetic constraints, filler-induced barriers, and structural discontinuities limited the observed crystallinity increase in the samples from post-process annealing.

\subsection{Tensile testing}

\begin{figure}[htbp]
    \centering
    \includegraphics[width=0.95\textwidth]{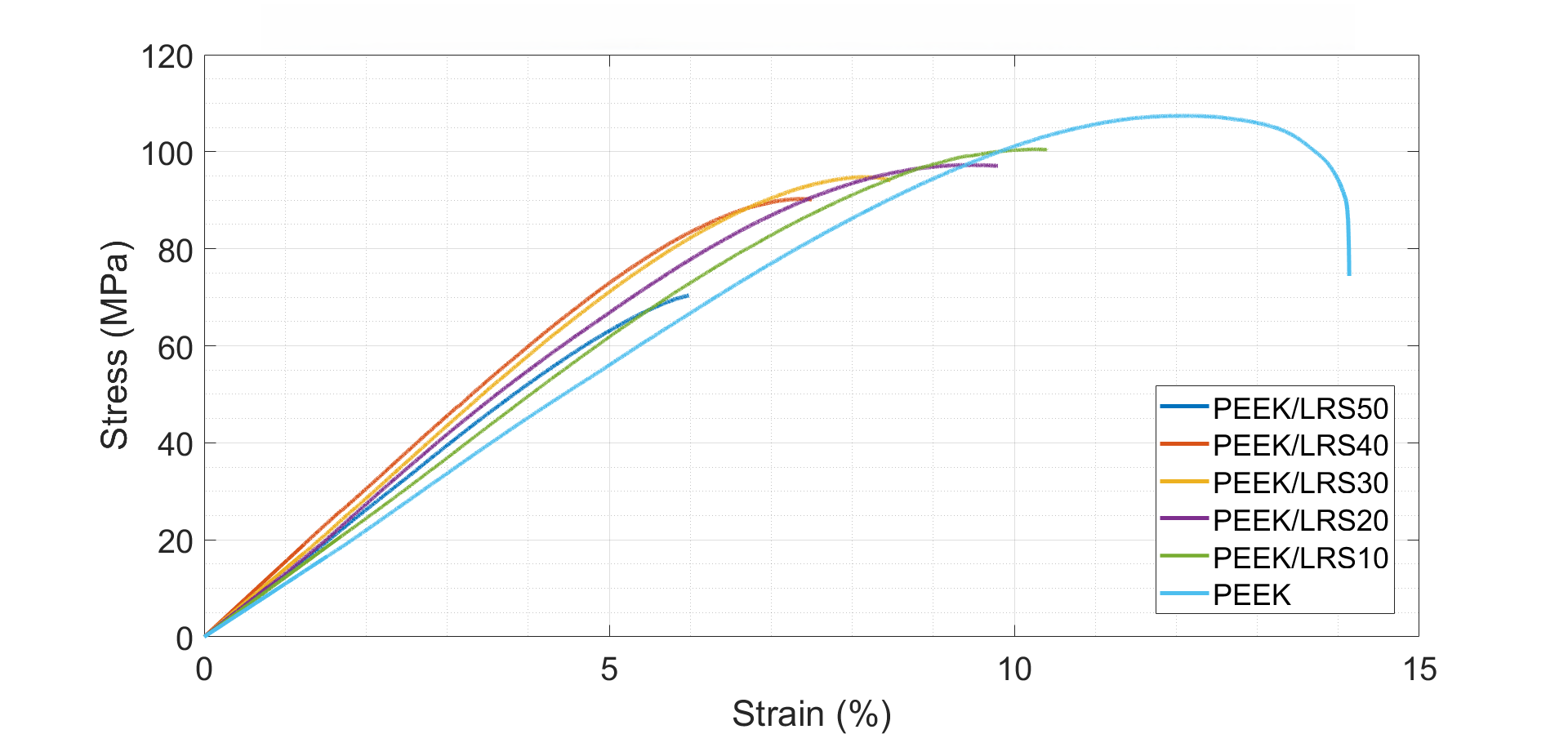}
    \caption{Engineering stress-strain curves of the as-printed samples fabricated in this study, including neat PEEK and PEEK/LRS composites with 10–50~wt\% lunar regolith simulant, showing the curves corresponding to median ultimate tensile strength values.}
    \label{tencom}
\end{figure}

Figure~\ref{tencom} presents the engineering stress-strain curves for the as-printed samples fabricated in the current work, including neat PEEK and PEEK/LRS composites containing 10–50\,wt\% lunar regolith simulant. As-printed \textbf{neat PEEK} exhibited the highest tensile performance among all samples, with an ultimate tensile strength (UTS) of 107.4\,MPa and elongation at break of 14.1\,\%. The material demonstrated ductile fracture behavior with a visible necking phenomenon prior to failure. When compared to the previous results reported by Azami et al.~\cite{azami2025enhancing}, this corresponds to a 17.3\,\% increase in UTS and a 30.1\,\% reduction in elongation at break. Given that the porosity levels in both studies remained nearly unchanged, these mechanical differences are attributed primarily to the increased crystallinity achieved in the current study. This enhancement is likely due to the use of a higher nozzle temperature combined with the activation of both DAS tuyeres during printing. These conditions promote better thermal management and improved interbead fusion. The Young's modulus for neat PEEK in the current work was 1,094.0\,MPa, reflecting a 19.3\,\% increase relative to the value reported in Azami et al \cite{azami2025enhancing}.

The addition of regolith (up to 40~wt\%) resulted in a gradual reduction in both elongation at break and ultimate tensile strength (UTS), while stiffness increased with filler content. The as-printed \textbf{PEEK/LRS10} sample exhibited an elongation at break of 10.4\,\%, a UTS of 100.5\,MPa, and a Young’s modulus of 1,225.3\,MPa, corresponding to a 26.2\,\% reduction in elongation, a 6.4\,\% decrease in UTS, and a 12.0\,\% increase in stiffness relative to neat PEEK. For as-printed \textbf{PEEK/LRS20}, the corresponding values were 9.8\,\%, 97.2\,MPa, and 1,307.9\,MPa, representing a 30.5\,\% reduction in elongation, a 9.5\,\% reduction in UTS, and a 19.6\,\% increase in elastic modulus relative to neat PEEK. The trend continued with as-printed \textbf{PEEK/LRS30}, which showed values of 8.5\,\%, 94.8\,MPa, and 1,402.6\,MPa, corresponding to a 39.7\,\% reduction in elongation, an 11.7\,\% decrease in UTS, and a 28.2\,\% increase in elastic modulus relative to neat PEEK. As-printed \textbf{PEEK/LRS40} exhibited elongation, UTS, and modulus values of 7.5\,\%, 90.3\,MPa, and 1,544.9\,MPa, respectively, representing a 46.8\,\% reduction in elongation, a 15.9\,\% decrease in UTS, and a 41.2\,\% increase in stiffness compared to neat PEEK.

A direct comparison with the as-printed PEEK/LRS30 data from Azami et al.~\cite{azami2025enhancing} further highlights the superiority of the current process. In the prior work, PEEK/LRS30 exhibited a tensile strength of only 67.1\,MPa, 7.9\,\% elongation, and a Young's modulus of 1,152.1\,MPa. In contrast, the present results show improvements of 41.3\,\% in UTS, 7.6\,\% in elongation, and 21.7\,\% in elastic modulus over the prior results. These improvements are attributed to higher print quality, diminished porosity, and stronger interbead bonding achieved through the refinements of the process parameters described earlier.

However, when the filler content was increased to 50\,wt\%, the mechanical performance began to deteriorate. The as-printed \textbf{PEEK/LRS50} composite showed a UTS of 70.4\,MPa, elongation of 6.0\,\%, and Young’s modulus of 1,268.2\,MPa. This represents a 22.0\,\% decrease in UTS, a 20.0\,\% reduction in elongation, and a 17.9\,\% decrease in stiffness compared to the PEEK/LRS40 sample. Interestingly, the modulus dropped below the value observed for PEEK/LRS30, indicating a deviation from the earlier increasing trend. This behavior is likely the result of excessive melt viscosity at such high filler concentrations, which severely restricted polymer chain mobility and extrusion flow, leading to higher porosity and weaker interbead bonding.

All PEEK/LRS composites exhibited brittle fracture behavior, characterized by the absence of necking. The mechanical properties follow a consistent trend with increasing regolith content: for every additional 10\,wt\% of filler, the UTS decreased by approximately 2.5–6.4\,\%, elongation at break dropped by less than 12\,\%, and Young’s modulus increased by 6.7–12.0\,\%. Between neat PEEK and PEEK/LRS10, for instance, there was a 26.2\,\% decrease in elongation, while the modulus increased by 12.0\,\%.

The observed mechanical response is governed by two competing mechanisms. On one hand, the presence of submicron- and nanoscale regolith particles may act as nucleating agents during solidification, enhancing crystallinity and contributing to improved stiffness and potentially strength \cite{abraham2009mechanical, wu2020recent}. On the other hand, increased regolith content raises the melt viscosity of the composite, making extrusion more difficult and increasing the risk of pore formation \cite{chuang2015additive}. It also limits chain mobility during annealing by acting as a barrier to polymer chain movement, thereby hindering crystallization. While the potential reinforcement effects of the nanoscale particles are beneficial, the increase in porosity due to poorer flow dominates, leading to reduced tensile strength despite gains in elastic modulus.

One final observation concerns the fracture surface morphology. The fracture surface of pure PEEK shows a distinct layered structure. With the addition of 10\,wt\% regolith particles, this layered structure becomes nearly indistinguishable, and for composites with higher filler contents, it is completely absent. This suggests a progressive improvement in interlayer bonding, which may enhance Z-direction mechanical properties. The improvement is likely due to the higher thermal conductivity of the regolith-filled composites, which promotes more uniform thermal distribution and better fusion between layers during the printing process.

\begin{figure}[htbp]
    \centering

    % Row: (a) UTS and (b) Young's Modulus
    \begin{subfigure}[b]{0.9\textwidth}
        \centering
        \includegraphics[width=\linewidth]{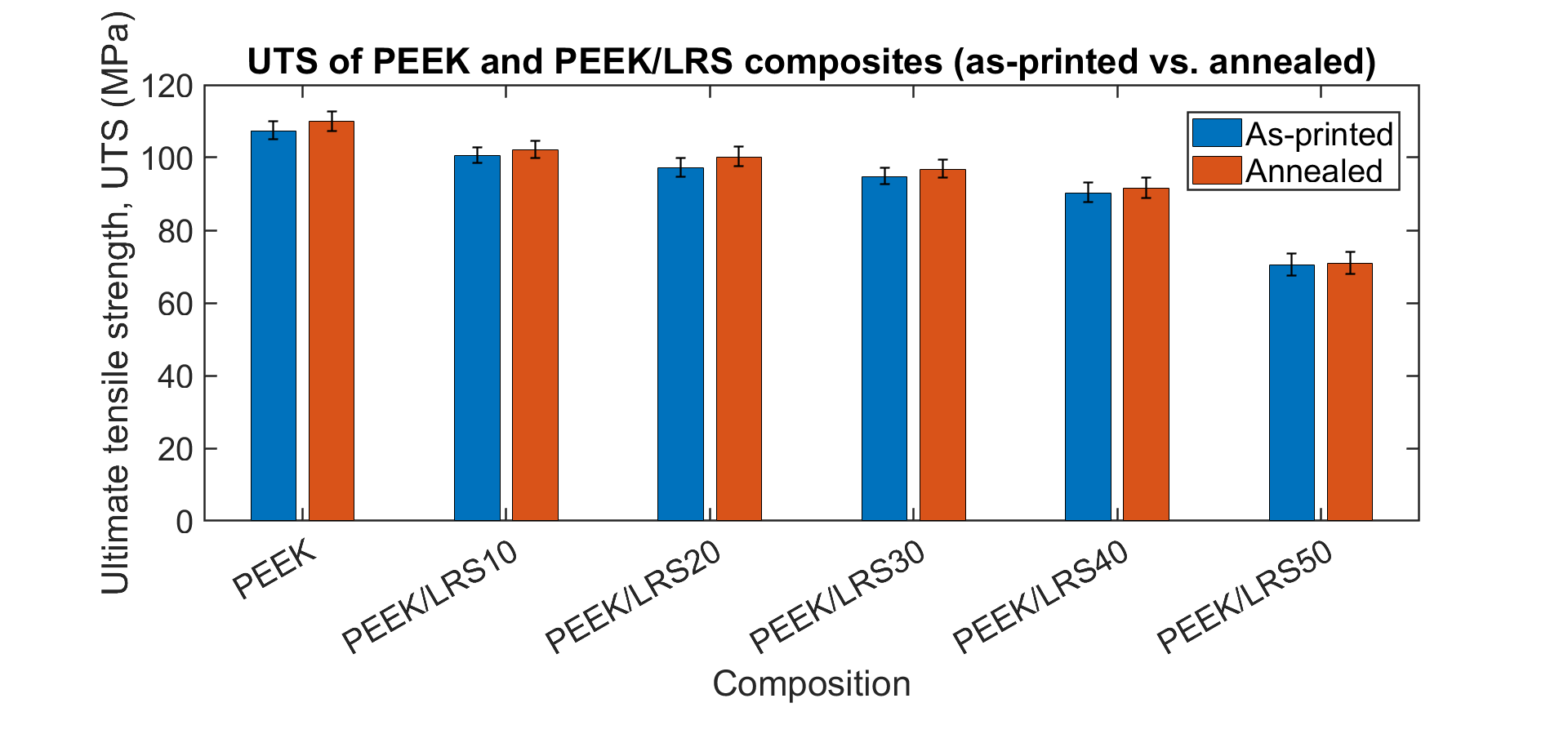}
        \caption{Ultimate tensile strength (UTS) of as-printed and annealed samples.}
        \label{fig:mech_UTS}
    \end{subfigure}
    \hfill
    \begin{subfigure}[b]{0.9\textwidth}
        \centering
        \includegraphics[width=\linewidth]{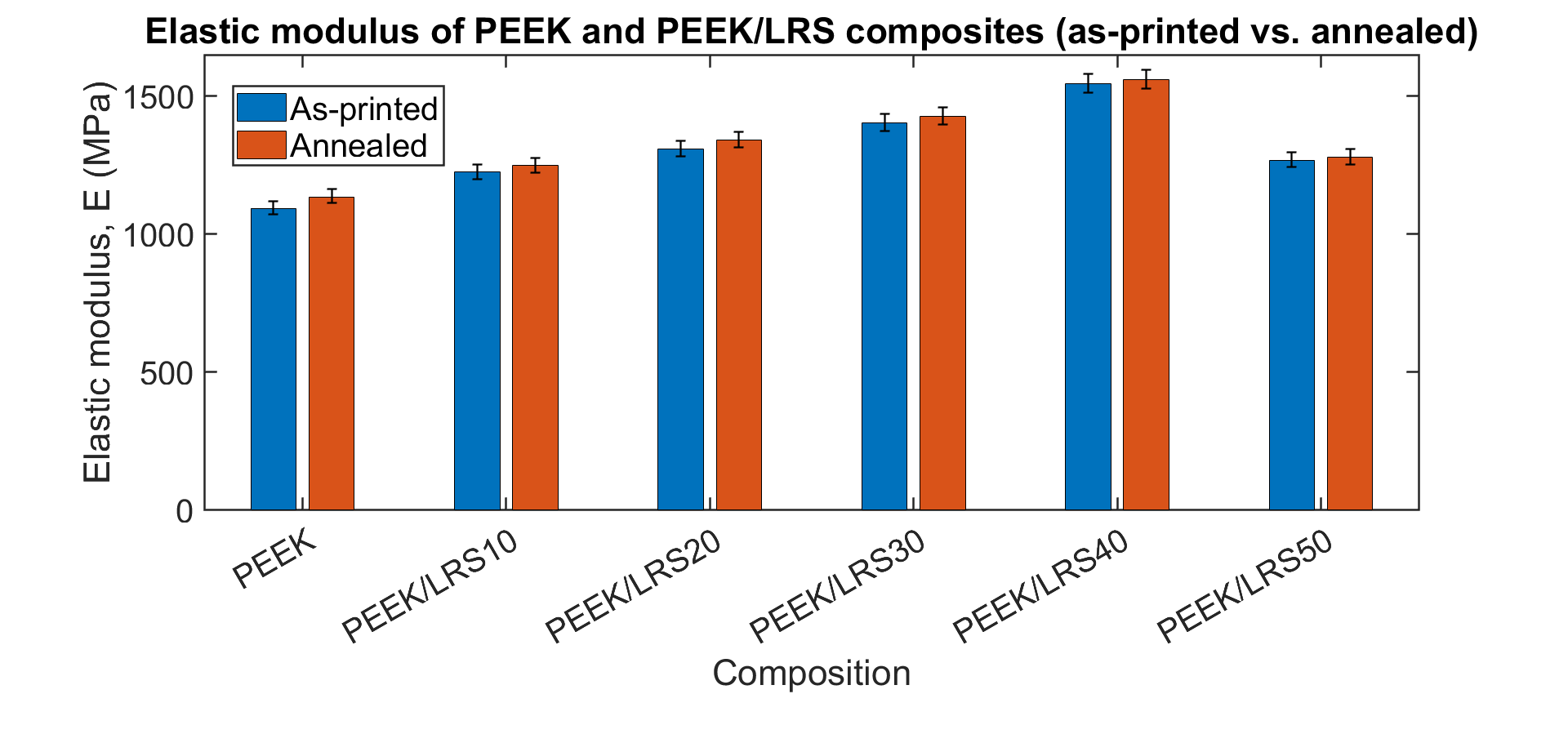}
        \caption{Young's modulus (E) of as-printed and annealed samples.}
        \label{fig:mech_E}
    \end{subfigure}

    \caption{Comparison of mechanical properties, including (a) UTS and (b) Young’s modulus, of as-printed versus annealed PEEK and PEEK/LRS composites.}
    \label{fig:UTS_E_comparison}
\end{figure}

As shown in Figure~\ref{fig:UTS_E_comparison}, following annealing, all compositions demonstrated modest yet consistent enhancements in mechanical performance, particularly in stiffness. For \textbf{neat PEEK}, the ultimate tensile strength (UTS) increased from 107.4 to 109.9\,MPa, representing a 2.3\% improvement, while the Young’s modulus rose from 1,094.0 to 1,135.3\,MPa. This corresponds to a 3.8\% increase, which is a larger effect than that observed on the UTS. It should be noted that the annealed PEEK sample retained a layered structure similar to that of the as-printed sample, indicating the limited effectiveness of the annealing process in altering the microstructural morphology.

In the case of \textbf{PEEK/LRS10}, the UTS increased from 100.5 to 102.1\,MPa (+1.6\,\%), and the modulus improved from 1\,225.3 to 1\,248.1\,MPa (+1.9\,\%). For \textbf{PEEK/LRS20}, the UTS rose to 100.2\,MPa (+3.1\,\%) and the modulus to 1\,340.5\,MPa (+2.5\,\%). \textbf{PEEK/LRS30} experienced more modest gains, with a UTS of 96.8\,MPa (+2.1\,\%) and a modulus of 1\,426.0\,MPa (+1.7\,\%). These enhancements are primarily attributed to the reduction in processing-induced porosity during the annealing step.

The enhancements were even less pronounced for higher filler loadings. \textbf{PEEK/LRS40} exhibited a UTS of 91.5\,MPa (+1.3\,\%) and a modulus of 1\,560.7\,MPa (+1.0\,\%), while \textbf{PEEK/LRS50} showed the smallest improvements, reaching 70.9\,MPa in UTS (+0.7\,\%) and 1\,278.2\,MPa in modulus (+0.8\,\%). As anticipated from the observed changes in density and crystallinity after annealing, the mechanical benefits become limited at higher filler contents. This is primarily due to the increased prevalence of structural defects and disrupted matrix continuity, which hinder chain mobility during annealing and collectively restrict further improvements in mechanical performance.

\iffalse
\begin{table}[ht]
\caption{Mechanical characteristics of as-printed PEEK/LRS composites and changes after a 300 \textdegree{}C anneal (\(2\;\text{h}\), N₂).  
Arrows denote the predicted post-anneal values; percentages in parentheses are the expected relative changes.}
\label{tab:mech-anneal}
\centering
\footnotesize
\setlength{\tabcolsep}{5pt}
\small
\begin{tabular}{l c c c}
\hline
\textbf{Regolith} &
\textbf{UTS} & \textbf{Elongation at break} & \textbf{Young’s modulus} \\
\textbf{content (wt\%)} &
(MPa) & (\%) & (MPa) \\
\hline
0  & 107.4 $\;\rightarrow\;$ 113\,(+5)  & 14.1 $\;\rightarrow\;$ 12.7\,(–10) & 1094 $\;\rightarrow\;$ 1200\,(+10) \\
10 & 100.5 $\;\rightarrow\;$ 105\,(+4)  & 10.4 $\;\rightarrow\;$  9.9\,(–5) & 1225 $\;\rightarrow\;$ 1300\,(+6)  \\
20 &  97.2 $\;\rightarrow\;$ 102\,(+5)  &  9.8 $\;\rightarrow\;$  9.2\,(–6) & 1308 $\;\rightarrow\;$ 1400\,(+7)  \\
30 &  94.8 $\;\rightarrow\;$  98\,(+3)  &  8.5 $\;\rightarrow\;$  8.2\,(–3) & 1403 $\;\rightarrow\;$ 1460\,(+4)  \\
40 &  90.3 $\;\rightarrow\;$  92\,(+2)  &  7.5 $\;\rightarrow\;$  7.3\,(–2) & 1545 $\;\rightarrow\;$ 1580\,(+2)  \\
50 &  70.4 $\;\rightarrow\;$  71\,(+1)  &  6.0 $\;\rightarrow\;$  5.9\,(–1) & 1268 $\;\rightarrow\;$ 1280\,(+1)  \\
\hline
\end{tabular}
\normalsize
\end{table}
\fi

\subsection{Microstructural analysis}

\subsubsection{Scanning electron microscopy (SEM)}

Backscattered electron (BSE) SEM micrographs of polished cross-sections are obtained for filaments and the as-printed samples with median UTS from the second phase. These micrographs support the reduction in mechanical performance observed in tensile testing with increasing regolith particle concentration. As the regolith content increases, a corresponding rise in defect incidence is observed.

\begin{figure}[htbp]
    \centering
    % Subfigure (a): Neat PEEK Filament
    \begin{subfigure}[b]{0.48\textwidth}
        \centering
        \includegraphics[width=\linewidth]{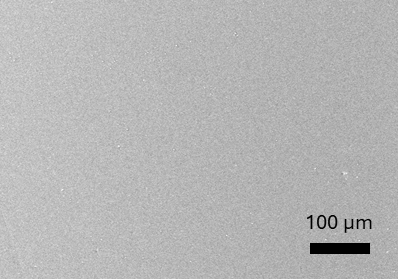} % Replace with your actual file name
        \caption{Neat PEEK filament.}
        \label{fig:peek_filament}
    \end{subfigure}
    \hfill
    % Subfigure (b): As-printed PEEK
    \begin{subfigure}[b]{0.48\textwidth}
        \centering
        \includegraphics[width=\linewidth]{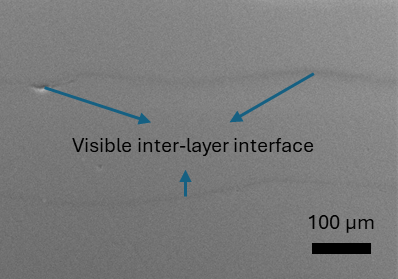} % Replace with your actual file name
        \caption{As-printed PEEK sample.}
        \label{fig:as_printed_peek}
    \end{subfigure}
    
    \caption{(a) Microstructure of the neat PEEK filament prior to printing. (b) Cross-section of the as-printed PEEK sample, where the interlayer boundary is clearly visible as a result of the layer-by-layer deposition process.}
    \label{fig:peek_microstructure}
\end{figure}

Figure~\ref{fig:peek_filament} presents the microstructure of the neat PEEK filament, which appears nearly defect-free. In contrast, Figure~\ref{fig:as_printed_peek} shows the cross-section of an as-printed PEEK sample. No significant defects are observed within individual layers; however, visible interlayer gaps are present, which are characteristic of the layer-by-layer deposition inherent to the MEX process. This morphological observation aligns with the fracture behavior seen during tensile testing. Fracture surfaces of neat PEEK specimens reveal a distinctly layer-controlled failure mode. Cracks typically initiate at the interlayer interfaces and propagate as planar segments oriented perpendicular to the loading direction. However, rather than forming a continuous break across the specimen width, the fracture surface often appears stepped, with adjacent layers failing at different positions along the gauge length. This staggered rupture pattern indicates non-uniform stress transfer across layers, resulting from limited polymer chain diffusion and insufficient interlayer entanglement during printing. Notably, despite achieving near-full bulk density (>99\%) as measured by the Archimedean method, the interlayer regions remain mechanically weak. Cracks tend to initiate and propagate along these interfaces under tensile loading or during post-fracture handling (e.g., cutting), underscoring the pronounced anisotropy inherent to MEX-processed PEEK.

\begin{figure}[htp]
    \centering

    % (a) PEEK/LRS10 – 100×
    \begin{subfigure}[b]{0.45\textwidth}
        \centering
        \includegraphics[width=\linewidth]{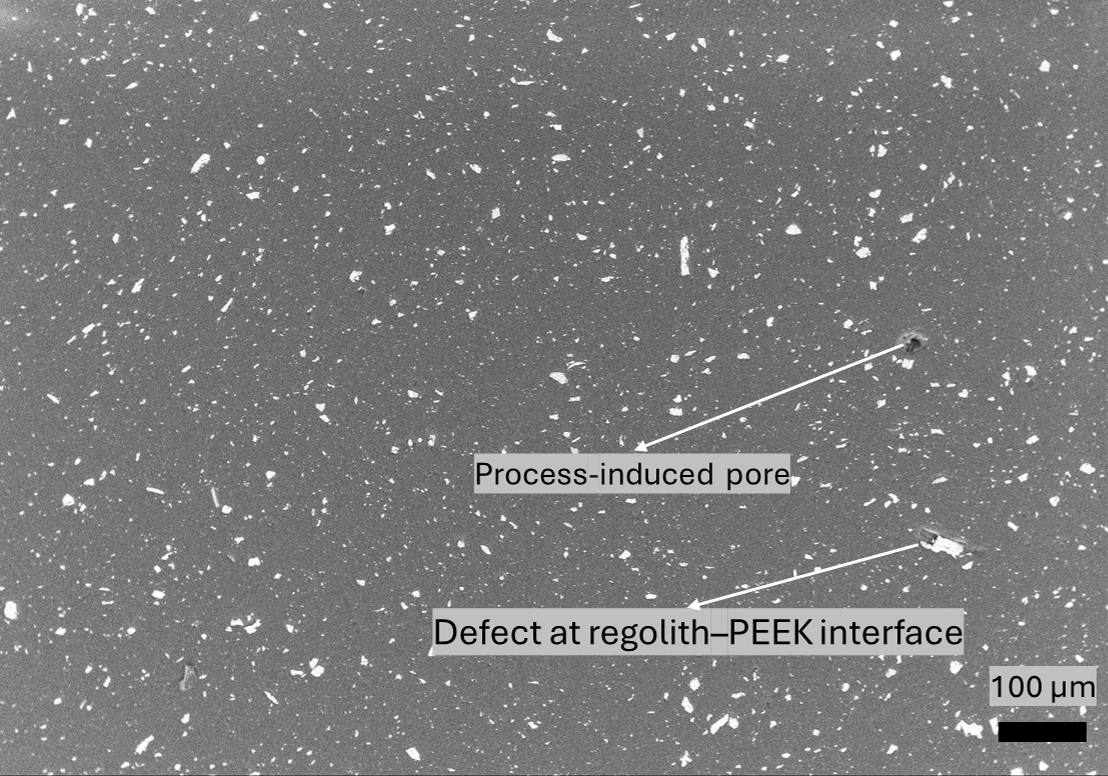} % Replace with actual filename
        \caption{PEEK/LRS10 (100×).}
        \label{fig:LRS10_100x}
    \end{subfigure}
    \hfill
    % (b) PEEK/LRS10 – 500×
    \begin{subfigure}[b]{0.45\textwidth}
        \centering
        \includegraphics[width=\linewidth]{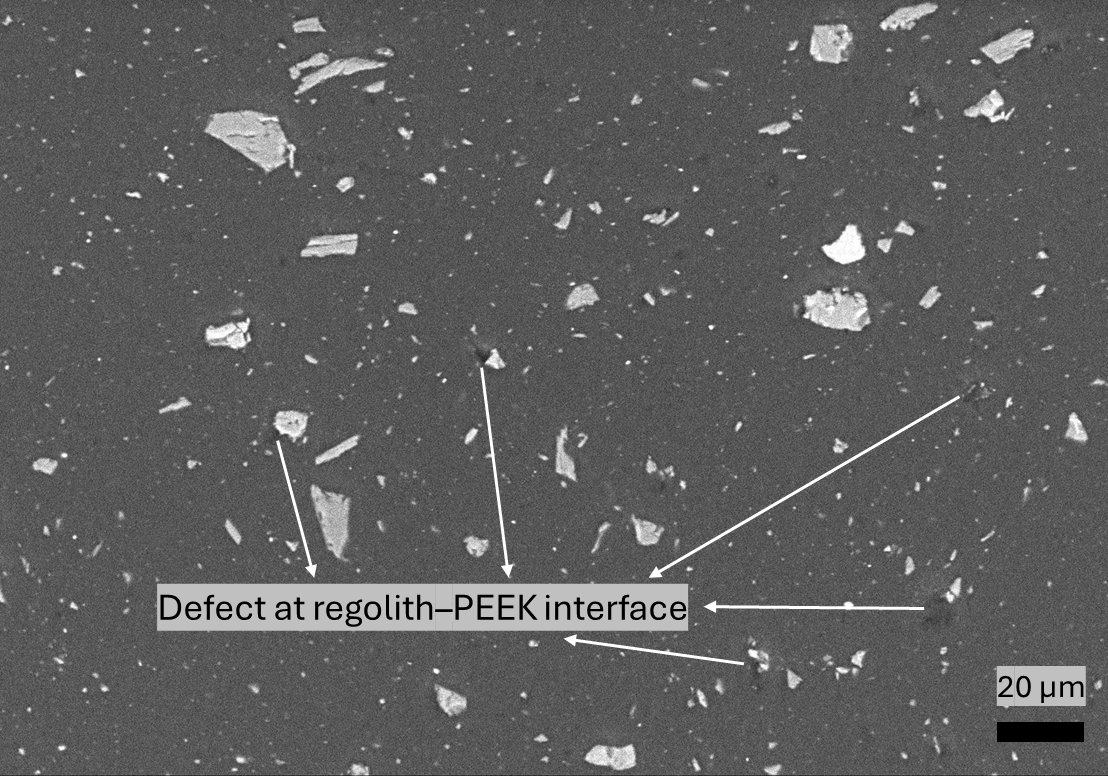} % Replace with actual filename
        \caption{PEEK/LRS10 (500×).}
        \label{fig:LRS10_500x}
    \end{subfigure}

    \vskip\baselineskip

    % (c) PEEK/LRS20 – 100×
    \begin{subfigure}[b]{0.45\textwidth}
        \centering
        \includegraphics[width=\linewidth]{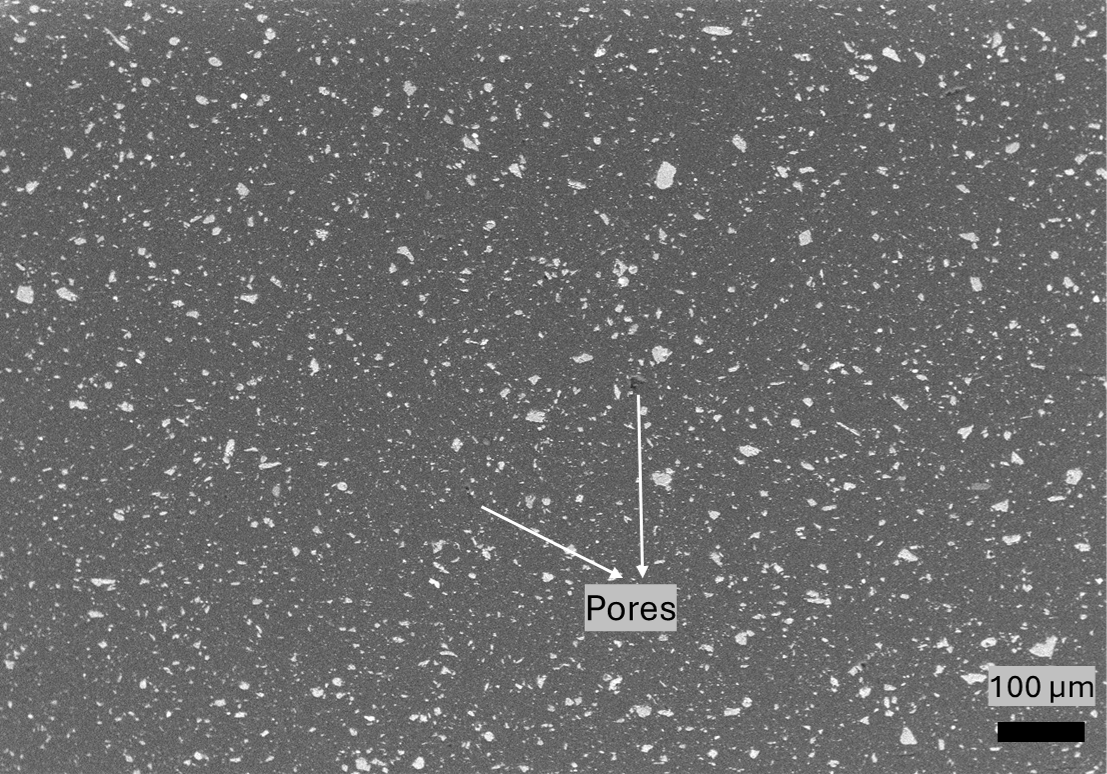} % Replace with actual filename
        \caption{PEEK/LRS20 (100×).}
        \label{fig:LRS20_100x}
    \end{subfigure}
    \hfill
    % (d) PEEK/LRS20 – 500×
    \begin{subfigure}[b]{0.45\textwidth}
        \centering
        \includegraphics[width=\linewidth]{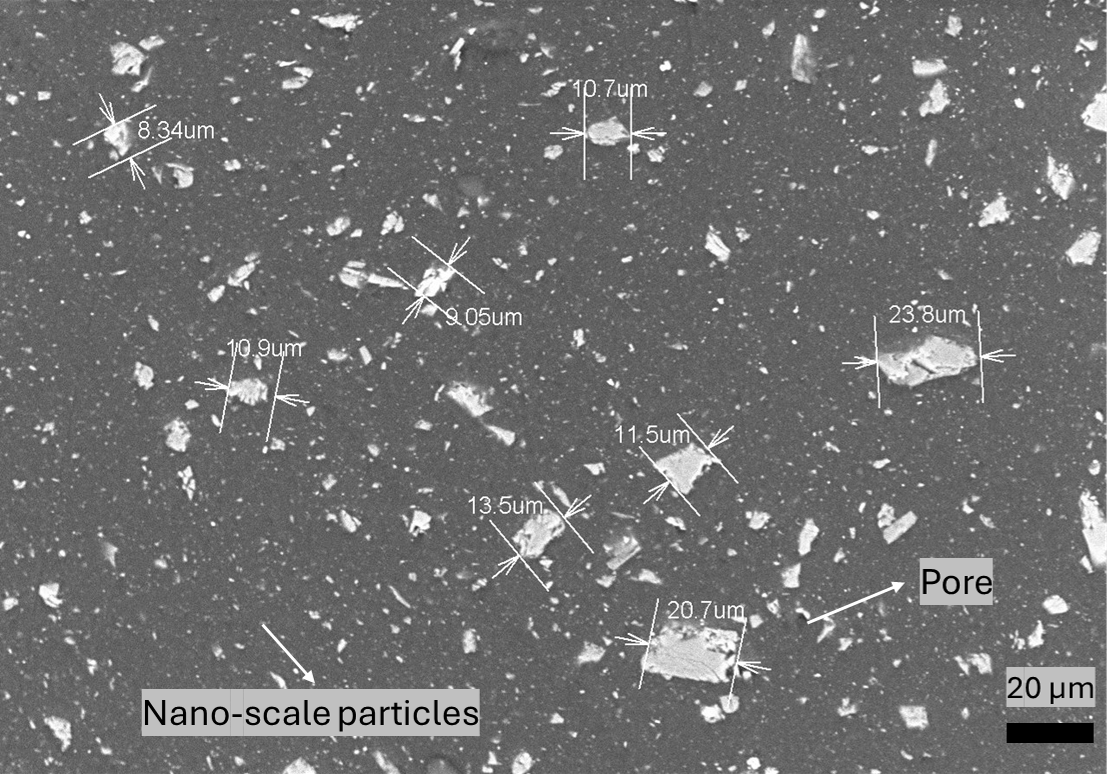} % Replace with actual filename
        \caption{PEEK/LRS20 (500×).}
        \label{fig:LRS20_500x}
    \end{subfigure}

    \caption{Backscattered SEM micrographs of polished cross-sections of as-printed PEEK/regolith composites. (a, b) PEEK/LRS10 and (c, d) PEEK/LRS20, shown at 100× and 500× magnifications.}
    \label{fig:LRS10_LRS20_SEM}
\end{figure}

The addition of LRS significantly alters the fracture behavior of the printed composites. Even at the lowest filler content investigated (PEEK/LRS10), the fracture path propagates uniformly across the specimen width and occurs nearly in the same plane perpendicular to the tensile loading direction, cutting through multiple printed layers rather than varying in position from layer to layer. SEM analysis of the fracture cross-sections (Figures~\ref{fig:LRS10_100x} and~\ref{fig:LRS10_500x}) reveals no discernible interlayer boundaries, indicating improved interlayer cohesion. This behavior suggests improved material integrity and interlayer bonding, even at modest regolith loadings. Two key mechanisms contribute to this altered failure mode. First, \textbf{particle-mediated stress redistribution}: the rigid ceramic particles promote mechanical bridging across adjacent rasters, enabling localized stress transfer between layers. As a result, crack fronts are forced to deflect or branch, increasing the fracture surface area and improving the effective interlayer fracture toughness. Second, \textbf{crystallization inhibition and thermal stress relaxation}: the presence of regolith particles during PEEK solidification disrupts spherulite growth, thereby reducing residual shrinkage stresses between layers. Although the thermal conductivity of the regolith is lower than that of metals, it remains significantly higher than that of polymers. This enhanced thermal conductivity facilitates better heat dissipation during printing, reduces thermal gradients, and minimizes residual stresses that typically arise from differential cooling.

The transition in fracture morphology is further supported by the observed increase in elastic modulus with increasing regolith content (up to 40~wt\%), despite reductions in ultimate tensile strength. These findings suggest that particle-induced stiffening and improved resistance to interlayer crack propagation can partially compensate for the brittleness introduced by the ceramic filler.

SEM micrographs of the as-printed PEEK/LRS10 sample (Figures~\ref{fig:LRS10_100x} and~\ref{fig:LRS10_500x}) reveal a uniform and randomly distributed dispersion of regolith particles within the PEEK matrix, demonstrating the effectiveness of the compounding and filament extrusion process. The observed process-induced pores are relatively small, typically less than 5\,µm in diameter, and are predominantly located near regolith particles. This spatial correlation suggests that pore formation may result from incomplete particle–matrix wetting or localized thermal shrinkage mismatch during solidification, leading to microvoid nucleation at the particle–polymer interface.

SEM micrographs of the PEEK/LRS20 composite (Figures~\ref{fig:LRS20_100x} and~\ref{fig:LRS20_500x}) reveal microstructural characteristics similar to those observed in the PEEK/LRS10 formulation. The pore population remains relatively low, with small and sparsely distributed voids consistent with the 2.4\,\% porosity measured via the Archimedean method. Notably, the increased regolith content does not lead to particle agglomeration; individual regolith grains remain well dispersed throughout the matrix, and their maximum observed diameter aligns with the nominal particle size distribution (PSD) of the LRS feedstock. At 500× magnification (Figure~\ref{fig:LRS20_500x}), sub-micron-scale fragments are also visible, likely corresponding to the fine tail of the PSD, further confirming the effectiveness of the mixing and extrusion process in achieving uniform dispersion without particle segregation.

\begin{figure}[htp]
    \centering

    % (c) Current work – 200×
    \begin{subfigure}[b]{0.45\textwidth}
        \centering
        \includegraphics[width=\linewidth]{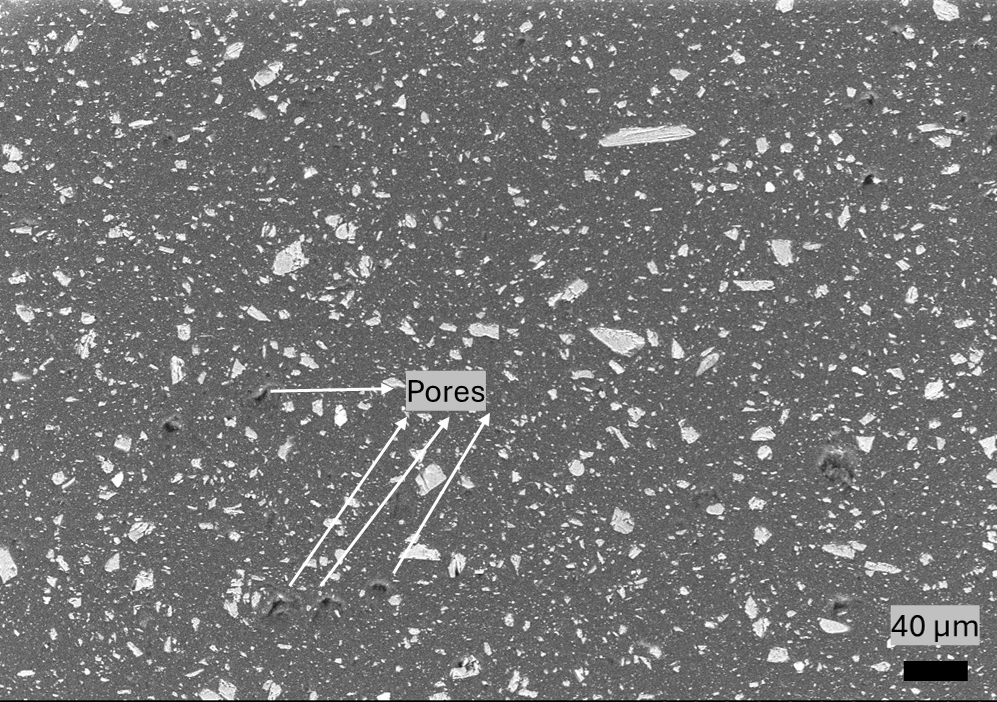} % Replace with actual filename
        \caption{PEEK/LRS30 (200×).}
        \label{fig:current_200x}
    \end{subfigure}
    \hfill
    % (d) Current work – 500×
    \begin{subfigure}[b]{0.45\textwidth}
        \centering
        \includegraphics[width=\linewidth]{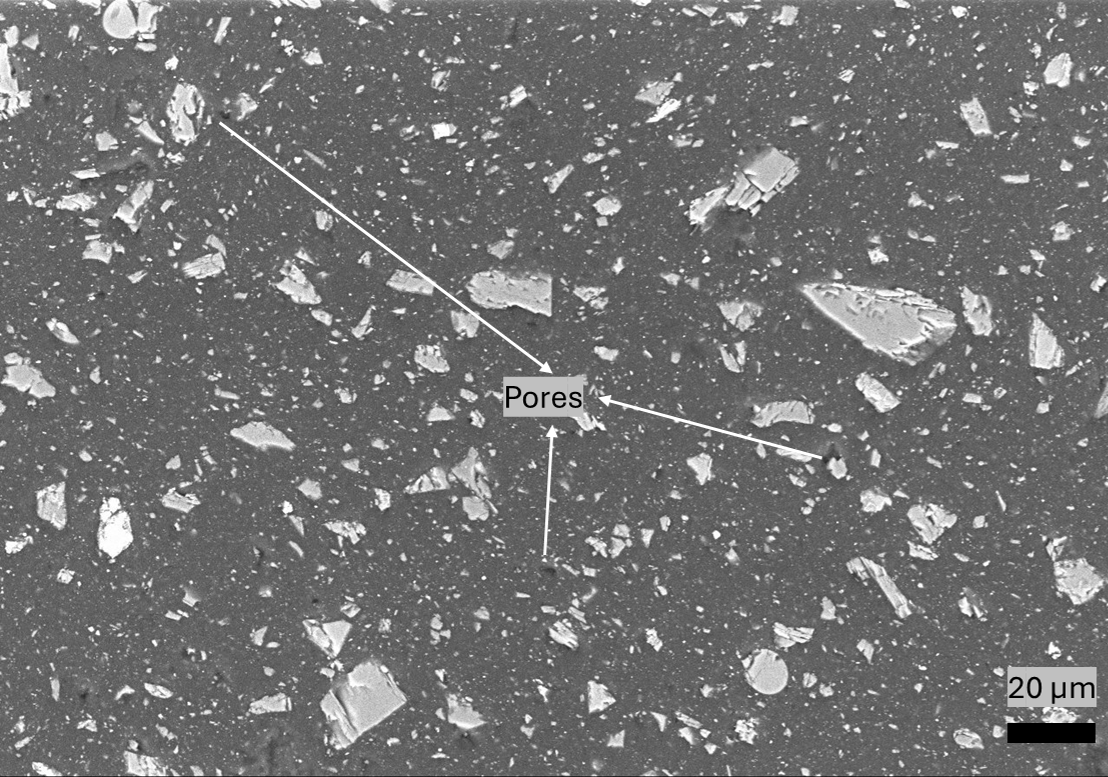} % Replace with actual filename
        \caption{PEEK/LRS30 (500×).}
        \label{fig:current_500x}
    \end{subfigure}

    \caption{Backscattered SEM micrographs showing polished cross-sections of as-printed PEEK/LRS30 samples fabricated in the present study, acquired at the same magnifications as those reported by Azami et al.~\cite{azami2025enhancing}. The microstructure observed in the current work exhibits visibly fewer and smaller defects, indicating a notable improvement in print quality.}
    \label{fig:MicrographyComparison}
\end{figure}

An increase in pore density is observed when transitioning from PEEK/LRS20 to PEEK/LRS30, as evident in Figures~\ref{fig:LRS20_500x} and~\ref{fig:current_500x}. This microstructural change aligns with a 1.4 percentage point decrease in relative density, as measured by the Archimedean method, and a corresponding 2.5\% reduction in ultimate tensile strength (UTS), from 97.2~MPa to 94.8~MPa. Despite the increased filler content, no particle agglomeration is observed, and the regolith remains well-dispersed throughout the matrix.

Furthermore, the SEM micrographs provide strong visual evidence of the enhanced mechanical performance achieved in this study compared to the results reported by Azami~\textit{et al.}~\cite{azami2025enhancing}. As shown in Figure~\ref{fig:MicrographyComparison}, the PEEK/LRS30 specimens fabricated in the present work exhibit noticeably fewer and smaller defects than those observed in the earlier study. This microstructural refinement corresponds to a substantial improvement in ultimate tensile strength (UTS), increasing from 67.1~MPa to 94.8~MPa, thereby highlighting the effectiveness of the optimized processing conditions employed in this work.

\begin{figure}[htp]
    \centering

    % Subfigure: PEEK/LRS40 filament
    \begin{subfigure}[b]{0.45\textwidth}
        \centering
        \includegraphics[width=\linewidth]{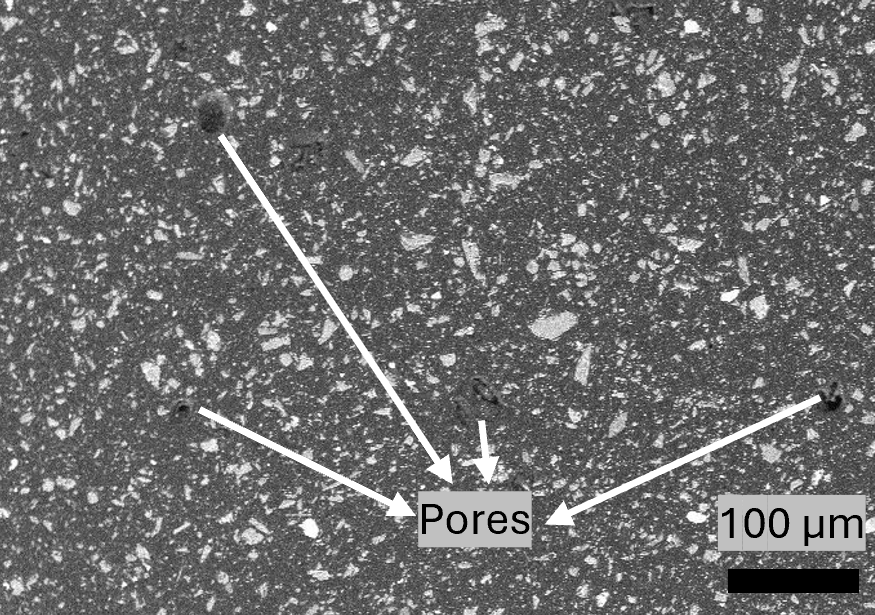} % Replace with actual filename
        \caption{PEEK/LRS40 filament.}
        \label{fig:40_fil}
    \end{subfigure}
    \hfill
    % Subfigure: PEEK/LRS50 filament
    \begin{subfigure}[b]{0.45\textwidth}
        \centering
        \includegraphics[width=\linewidth]{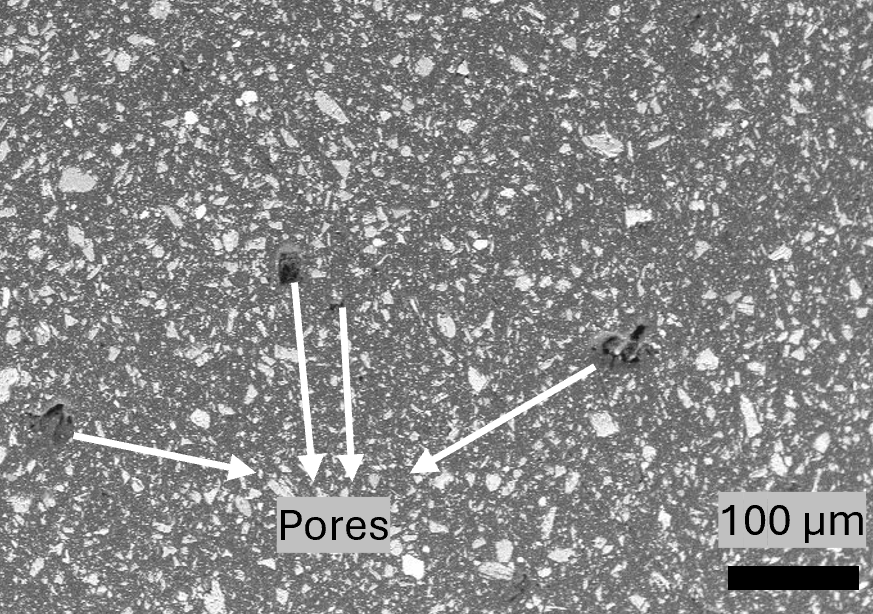} % Replace with actual filename
        \caption{PEEK/LRS50 filament.}
        \label{fig:50_fil}
    \end{subfigure}

    \caption{Backscattered SEM micrographs showing the cross-sectional morphology of extruded filaments: (a) PEEK/LRS40 and (b) PEEK/LRS50.}
    \label{fig:filament_SEM4050}
\end{figure}

As shown in Figure~\ref{fig:filament_SEM4050}, BSE–SEM micrographs of the PEEK/LRS40 and PEEK/LRS50 filaments reveal only minor differences in defect density and pore size between the two compositions. This observation is supported by Archimedean porosity measurements, which show just a 1.3 percentage point decrease in porosity when increasing the regolith content from 40 to 50~wt\%. However, the PEEK/LRS50 filament exhibits a noticeably rougher surface, likely due to the higher melt viscosity of the composite at elevated filler loadings. The increased regolith content also makes the filament more brittle, consistent with its higher ceramic fraction.

\begin{figure}[htp]
    \centering
    % (a) PEEK/LRS40 at 100x
    \begin{subfigure}[b]{0.45\textwidth}
        \centering
        \includegraphics[width=\linewidth]{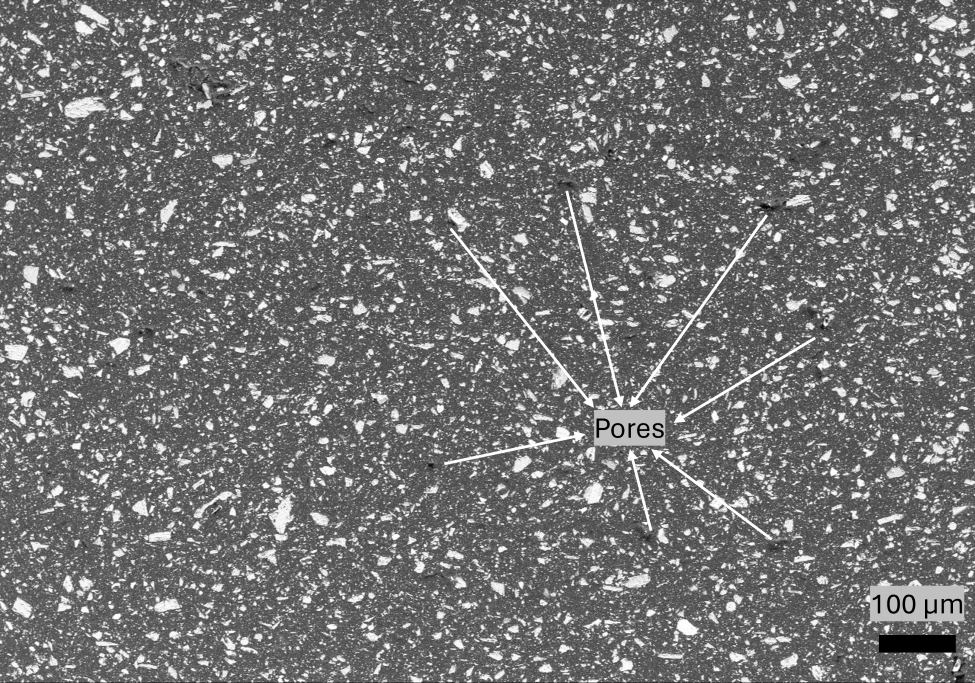} % Replace with actual filename
        \caption{PEEK/LRS40 as-printed (100×).}
        \label{fig:40_100x}
    \end{subfigure}
    \hfill
    % (b) PEEK/LRS40 at 500x
    \begin{subfigure}[b]{0.45\textwidth}
        \centering
        \includegraphics[width=\linewidth]{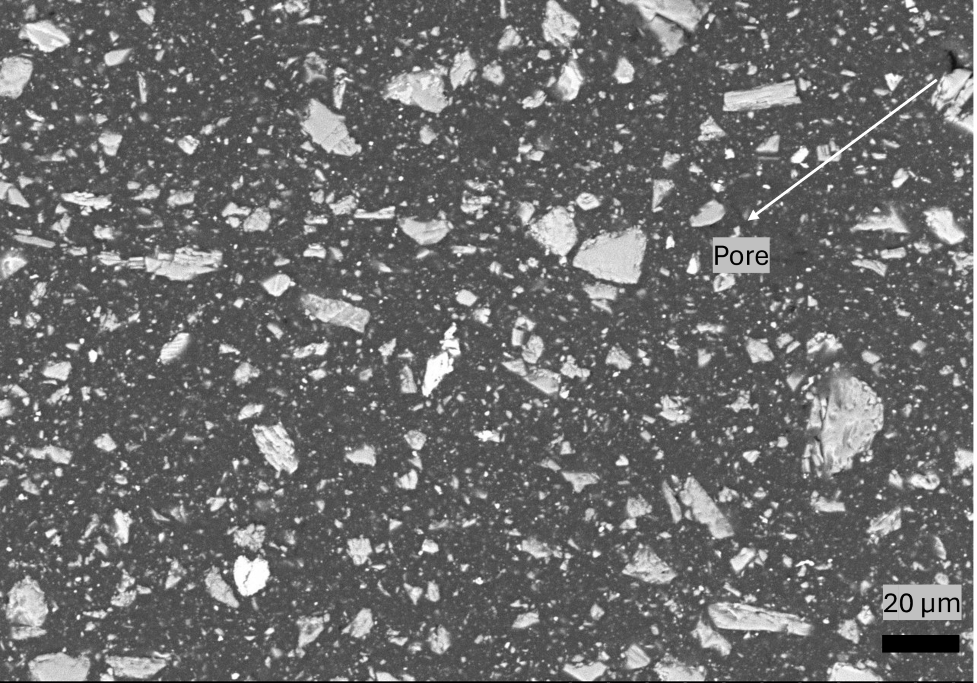} % Replace with actual filename
        \caption{PEEK/LRS40 as-printed (500×).}
        \label{fig:40_500x}
    \end{subfigure}

    \vskip\baselineskip

    % (c) PEEK/LRS50 at 100x
    \begin{subfigure}[b]{0.45\textwidth}
        \centering
        \includegraphics[width=\linewidth]{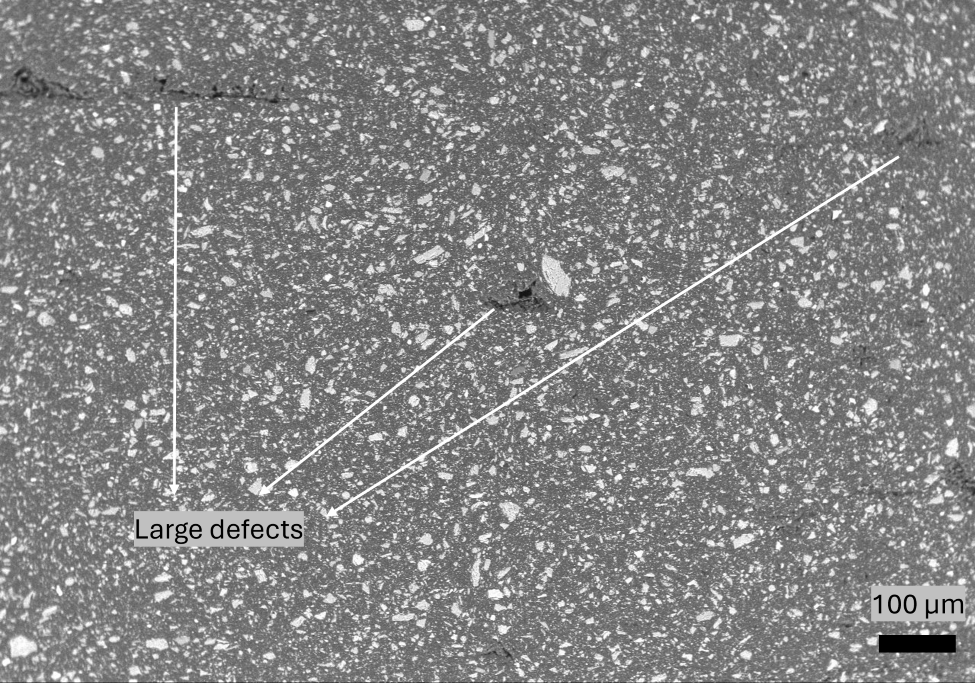} % Replace with actual filename
        \caption{PEEK/LRS50 as-printed (100×).}
        \label{fig:50_100x}
    \end{subfigure}
    \hfill
    % (d) PEEK/LRS50 at 500x
    \begin{subfigure}[b]{0.45\textwidth}
        \centering
        \includegraphics[width=\linewidth]{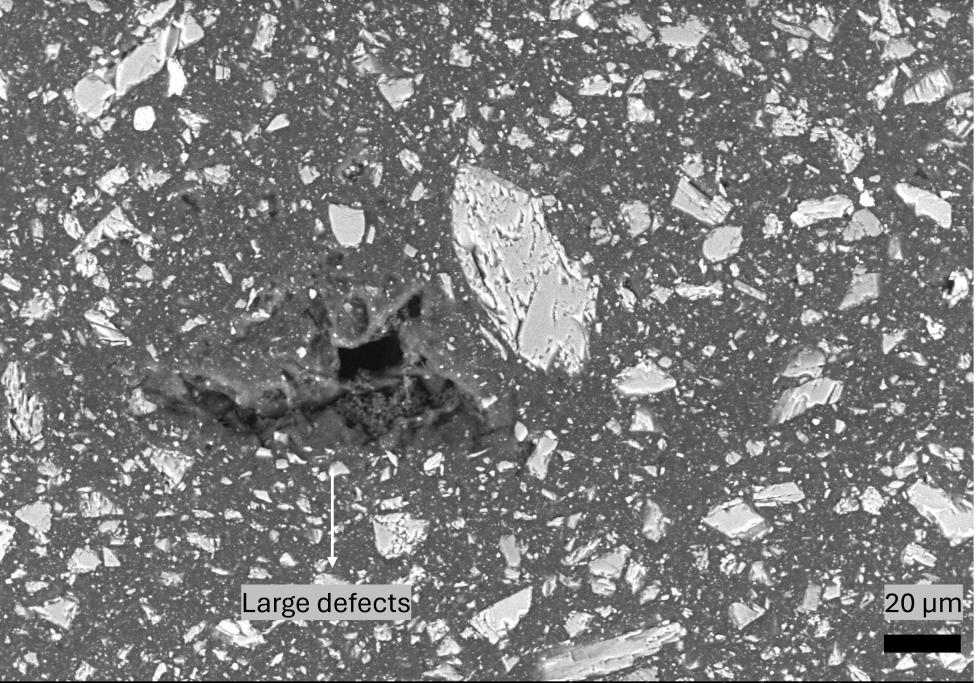} % Replace with actual filename
        \caption{PEEK/LRS50 as-printed (500×).}
        \label{fig:50_500x}
    \end{subfigure}

    \caption{Backscattered SEM micrographs showing polished cross-sections of as-printed samples: (a, b) PEEK/LRS40 and (c, d) PEEK/LRS50 at 100× and 500× magnifications, respectively.}
    \label{fig:as_printed_SEM4050}
\end{figure}

Figures~\ref{fig:40_100x} and~\ref{fig:40_500x} present SEM micrographs of the as-printed PEEK/LRS40 sample. A gradual increase in both pore number and size is evident in samples ranging from neat PEEK to PEEK/LRS40, aligning well with the trends observed in porosity and mechanical performance based on Archimedean and tensile test results, respectively.

In contrast to the relatively small changes observed in the filaments (Figure~\ref{fig:filament_SEM4050}), SEM images of the corresponding as-printed samples (Figure~\ref{fig:as_printed_SEM4050}) display a pronounced increase in both the density and size of defects when the regolith content is raised from 40 to 50~wt\%. This microstructural deterioration is consistent with the 2.6 percentage point increase in porosity observed via the Archimedean method and corresponds with the measured decline in mechanical properties at higher regolith loadings.

All SEM analyses in this study were conducted on the fracture surfaces of the extruded filaments and as-printed specimens. Although a limited set of annealing experiments was conducted to evaluate crystallinity gains and defect closure, the primary focus of this work remains on as-printed properties. This choice is deliberate and grounded in practical constraints specific to space-based ISRU manufacturing.

First, annealing introduces an additional thermal cycle that runs counter to the ISRU objectives of minimizing operational complexity, energy consumption, and processing time. Transporting an oven to the Moon incurs significant cost and logistical challenges. Printing 1~kg of PEEK typically consumes 1--5~kWh depending on printing parameters, whereas heating and maintaining a well-insulated 30~$\times$~30~$\times$~30~cm oven at 300\textdegree{}C for a 2-hour annealing cycle requires approximately 5~kWh, placing the annealing energy demand within the same range or potentially exceeding the energy required for printing alone. For off-Earth manufacturing systems, where power and time are highly constrained, this extra step, along with the necessary hardware and energy expenditure, represents an inefficient overhead. From a system-level perspective, the added equipment, extended processing time, and substantial energy demands are difficult to justify given the relatively marginal mechanical property improvements achieved through annealing.

Second, repeated thermal exposure can negatively affect the long-term durability and reusability of thermoplastic components. PEEK is susceptible to molecular chain scission and oxidative degradation under extended or repeated high-temperature cycles~\cite{day1988thermal, gaitanelis2023multi}. For ISRU applications, where components may be repurposed, repaired, or recycled across multiple mission cycles, limiting cumulative thermal history helps preserve performance and extend material service life.

Third, and most critically, the mechanical and densification benefits of annealing diminish substantially at higher regolith contents. As detailed in Table~\ref{density}, annealing almost fully eliminated residual porosity in neat PEEK and composites up to 20\,wt\% regolith. However, for 30–50\,wt\% formulations, the relative density gains were minimal (typically less than 1\,\%) and produced negligible improvements in tensile strength. These high regolith loadings are the most relevant for ISRU, as they most significantly reduce the amount of imported PEEK required. For example, composites with 50\,wt\% regolith content halve the mass of polymer needed, which is a key advantage in logistics-constrained environments. At these concentrations, porosity and interlayer defects become the dominant limiting factors while annealing offers little remediation. Conversely, for regolith contents up to 30–40\,wt\%, the as-printed quality is already sufficient for many aerospace structures, with only slightly lower mechanical performance compared to post-process annealed parts.

\subsubsection{Elemental composition characterization}

\begin{figure}[htbp]
    \centering
    \includegraphics[width=0.5\textwidth]{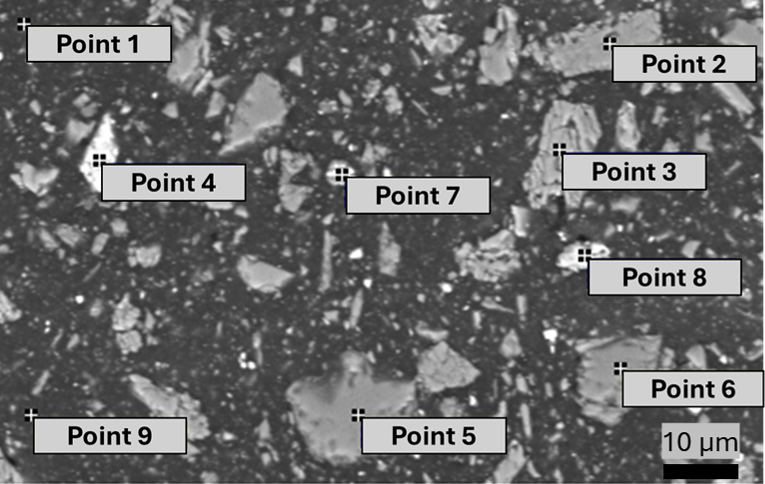}
    \caption{Backscattered SEM micrograph of the PEEK/LRS50 sample showing selected EDS points. Points 1 and 9 correspond to the PEEK matrix, while the remaining points represent various regolith particles.}
    \label{fig:EDS_points}
\end{figure}

Energy-dispersive X-ray spectroscopy (EDS) was performed at the nine labelled sites shown in Figure~\ref{fig:EDS_points}, and the corresponding elemental compositions are summarized in Table~\ref{tab:EDS_composition}. Elemental distribution maps further illustrate, in line with the SEM data, that the regolith particles are dispersed randomly and uniformly throughout the matrix, highlighting the consistency of the material's microstructure.

\textbf{Matrix locations (Points 1 and 9).} These regions exhibited high carbon (74--79~wt\%) and oxygen (18--24~wt\%) content, with only trace amounts of metals. This composition is consistent with the polymer matrix, confirming these points are located within the PEEK phase. Minor elements such as Si, Na, and Fe were detected at levels below 1.5~wt\%, likely attributable to surface contamination or polishing debris.

\textbf{Mg--Si-rich silicates (Points 2 and 3).} These inclusions were dominated by oxygen (60--63~wt\%), magnesium (13--15~wt\%), and silicon (18--20~wt\%), along with minor iron (2--3~wt\%). Such a composition suggests the presence of pyroxene or olivine fragments, commonly found in regolith simulants.

\textbf{Na--Al--Ca feldspathic particles (Points 4 and 5).} These sites contained oxygen concentrations around 63~wt\%, along with notable amounts of Na (1.6--2.7~wt\%), Al (11--12~wt\%), Si (17--18~wt\%), and Ca (4--5~wt\%). The overall chemistry aligns with plagioclase-type feldspars ranging from sodic to calcic composition.

\textbf{Ti-bearing oxide inclusions (Points 6 and 8).} These regions were characterized by high oxygen content (73--75~wt\%) and elevated titanium (13--14~wt\%), along with iron (6--9~wt\%), and minor amounts of Mg, Si, and Mn. This composition is typical of ilmenite-like phases, frequently observed in lunar regolith.

\textbf{Fe-rich inclusion (Point 7).} This point revealed an exceptionally high iron content (48.7~wt\%), with reduced oxygen (35.6~wt\%), accompanied by Mg, Si, and minor Ti and Al. The data indicate a dense iron oxide such as magnetite.

In summary, the EDS results effectively distinguish the carbon-rich PEEK matrix from the oxide-based regolith particles, while also highlighting the chemical heterogeneity of the filler phase. The identified phases include Mg--Si silicates, feldspathic materials, ilmenite, and Fe-rich inclusions, all of which are representative of the complex oxide distribution expected in lunar regolith simulants. These findings have direct implications for the thermal, mechanical, and interfacial behavior of the composite.

\begin{table}[ht]
\caption{Elemental composition (wt\%) at the nine EDS acquisition points shown in Fig.~\ref{fig:EDS_points}.}
\label{tab:EDS_composition}
\centering
\footnotesize
\setlength{\tabcolsep}{4pt}
\small
\begin{tabular}{l c c c c c c c c c c}
\hline
\textbf{Point} & \textbf{C} & \textbf{O} & \textbf{Na} & \textbf{Mg} & \textbf{Al} & \textbf{Si} & \textbf{Ca} & \textbf{Ti} & \textbf{Mn} & \textbf{Fe} \\
\hline
Point 1 & 73.64 & 23.66 & 0.19 & 0.56 & 0.37 & 1.27 & 0.16 & 0.16 & --   & --   \\
Point 2 & --    & 63.44 & --   & 13.08& 0.94 & 18.28& 2.04  & --   & --   & 2.21 \\
Point 3 & --    & 60.36 & --  & 14.81& 1.02 & 20.21& 0.64   & --   & --   & 2.97 \\
Point 4 & --    & 62.94 & 1.64 & 0.48 & 12.12& 17.62& 5.20 & --   & --   & --   \\
Point 5 & --    & 63.62 & 2.73 & 0.53 & 11.24& 17.85& 4.04 & --   & --   & --   \\
Point 6 & -- & 72.82 & -- & 1.05  & 0.73 & 1.84 & 0.54 & 13.05& 0.74 & 9.24  \\
Point 7 & -- & 35.57 & -- & 5.48  & 1.57 & 6.76 & 0.82 & 1.10 & --  & 48.69\\
Point 8 & -- & 74.62 & -- & 1.64  & 0.91 & 1.95 & -- & 13.50 & 0.65 & 6.73  \\
Point 9 & 78.72 & 18.20 & -- & 1.22 & 0.25 & 1.39  & --   & --   & --   & 0.22 \\
\hline
\end{tabular}
\normalsize
\end{table}

\begin{figure}[htbp]
    \centering
    \includegraphics[width=\textwidth]{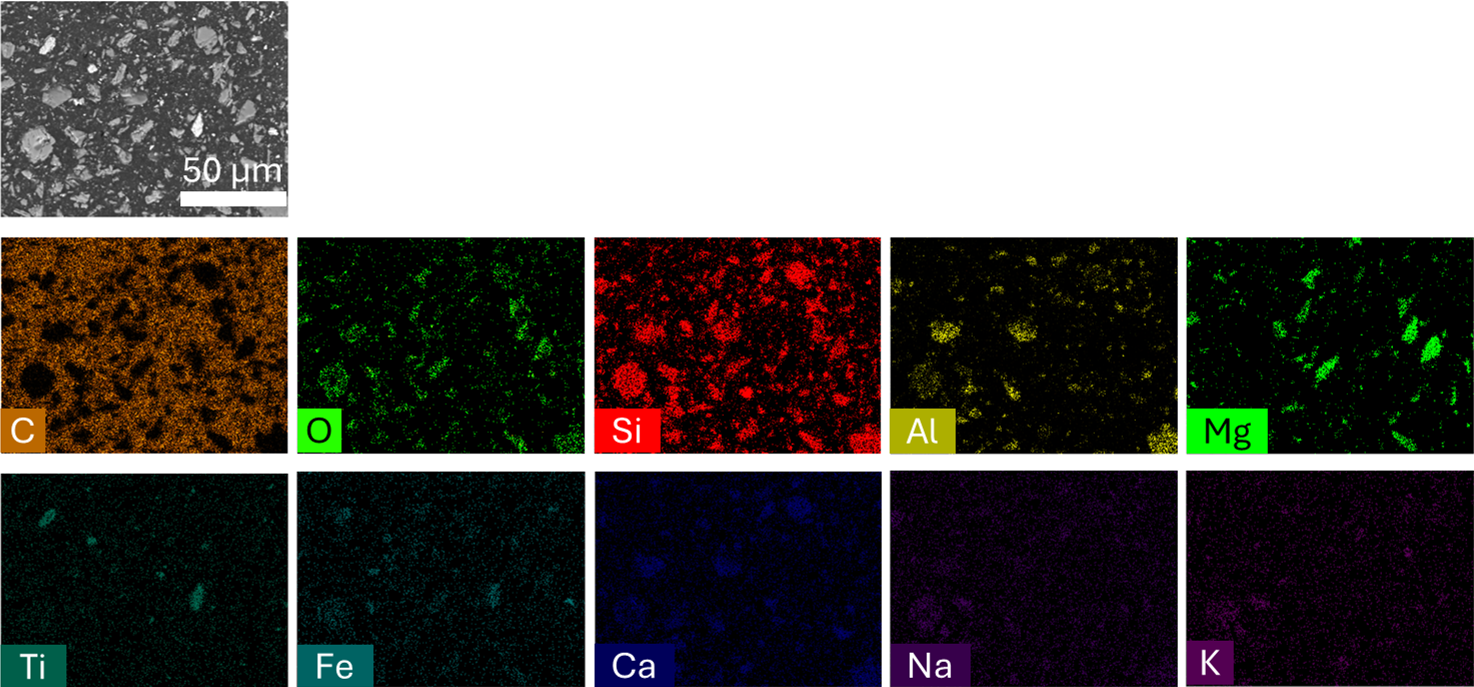}
    \caption{Backscattered electron SEM micrograph (a) and EDS X-ray maps of the as-printed PEEK/LRS50 composite acquired using C~$K_{\alpha}$, O-$K_{\alpha}$, Al-$K_{\alpha}$, Na-$K_{\alpha}$, Mg-$K_{\alpha}$, Si-$K_{\alpha}$, Ca-$K_{\alpha}$, Ti-$K_{\alpha}$, Fe-$K_{\alpha}$, and K-$K_{\alpha}$ radiations. The maps corroborate the point-analysis data, delineating the carbon-rich PEEK matrix from oxide-based regolith particles and highlighting the heterogeneous distribution of feldspathic, Mg–silicate, Ti–Fe oxide, and Fe-rich phases.}
    \label{fig:EDS_maps}
\end{figure}

Figure~\ref{fig:EDS_maps} presents a BSE SEM overview (upper-left) together with quantitative X-ray maps for the principal elements detected.  A clear two-phase architecture emerges:

\textbf{Carbon map.} A bright, continuous signal outlines the polymer phase, while angular and rounded voids correspond to the locations of regolith particles. This contrast confirms that carbon is almost entirely confined to the PEEK matrix.

\textbf{Oxygen map.} Oxygen is broadly distributed, with the highest intensity observed in the carbon-deficient regions, highlighting the oxide-rich nature of the filler particles. In contrast, the matrix shows a lower oxygen signal.

\textbf{Silicon and aluminum maps.} Silicon exhibits strong and widespread contrast across nearly all regolith inclusions, while aluminum is concentrated in a subset of these Si-bearing regions. Their co-localization is characteristic of feldspathic silicates (e.g., Na- or Ca-rich plagioclase), consistent with the earlier point analyses.

\textbf{Magnesium map.} Magnesium appears in discrete, elongated clusters that coincide with the Mg--Si-enriched domains identified in Table~\ref{tab:EDS_composition} (Points~2 and~3). These areas are interpreted as fragments of Mg-rich pyroxene or olivine.

\textbf{Titanium and iron maps.} Titanium shows a sparse distribution of bright pixels that often coincide with intense iron signals, indicating the presence of Ti--Fe oxides such as ilmenite. Several areas exhibit strong iron with negligible titanium, consistent with the Fe-rich particle observed at Point~7 in Figure~\ref{fig:EDS_points}.

\textbf{Calcium, sodium, and potassium maps.} Calcium is localized within a limited number of grains and commonly co-occurs with silicon and aluminum, suggesting the presence of calcic plagioclase. Sodium and potassium appear as weak, diffuse signals, broadly following the feldspathic regions. Their low intensity agrees with the quantitative point spectra.

To summarize, the elemental maps strongly corroborate the EDS point analysis: (i) the carbon-rich continuous phase is unambiguously identified as PEEK, and (ii) the embedded filler phase consists of a chemically heterogeneous mixture of feldspathic materials, Mg-silicates, Ti--Fe oxides, and isolated Fe-rich particles. This compositional diversity, clearly visualized in the X-ray maps, reflects the mineralogical complexity of the lunar regolith simulant and underscores the pronounced matrix--filler contrast that governs local thermal and mechanical behavior.

\section{Conclusion and future work}

This study demonstrates the material extrusion (MEX) additive manufacturing of polyether–ether–ketone (PEEK) and its composites containing 10–50~wt\% lunar regolith simulant (LRS), addressing key challenges in developing ISRU-compatible structural materials for lunar infrastructure. Filaments were produced via twin-screw extrusion, printed in a high-temperature chamber, and post-processed by annealing at 300~$^{\circ}$C. The resulting composites were systematically evaluated in terms of density, crystallinity, mechanical performance, and microstructural characteristics.

The main findings are as follows:

\begin{itemize}
    \item \textit{Processability:} All filament compositions achieved densities above 96\%. However, as-printed porosity rose from under 1\% in neat PEEK to 7.5\% at 50~wt\% LRS due to increased melt viscosity. A practical printability limit of ~40~wt\% LRS is identified for mechanically critical parts, while 50~wt\% remains viable for non-load-bearing or secondary structures. LRS incorporation and the use of a PEKK raft reduced delamination and warping, leading to improved dimensional accuracy and higher print success rates compared to pure PEEK.

    \item \textit{Thermal and mechanical performance:} Regolith addition increased matrix crystallinity from 17.4~\% to 20.5~\%, and enhanced the elastic modulus by 6–41~\%. Tensile strength declined steadily from 107~MPa for neat PEEK to 90~MPa at 40~wt\% LRS, followed by a sharper reduction to approximately 70~MPa at 50~wt\%.

    \item \textit{Post-processing response:} Furnace annealing improved both density and stiffness in composites containing up to 30~wt\% LRS. Above this threshold, gains were minimal, likely due to entrapped porosity, indicating the need for advanced post-processing methods at higher filler contents.

    \item \textit{Microstructural integrity:} Scanning electron microscopy (SEM) and energy-dispersive spectroscopy (EDS) revealed a carbon-rich PEEK matrix with uniformly dispersed regolith particles. The detected fillers included feldspathic, Mg-silicate, Ti--Fe oxide, and Fe-rich phases, confirming effective mixing and good phase contrast.
\end{itemize}

Overall, the results define a viable processing window for regolith-filled thermoplastic composites, with 40~wt\% LRS representing an important trade-off point balancing between mechanical performance and reduction in Earth-derived material. At 50~wt\%, the composite remains printable and structurally coherent, offering potential for non-critical or shielding applications where high stiffness is valued over strength.

Future work should focus on extending the printing process to simulated vacuum and low-gravity environments, exploring additional lunar regolith simulants and alternative high-performance polymers, and integrating finite element simulation to model material flow and stress evolution. Further studies are also needed to assess durability under lunar-relevant thermal cycling, radiation exposure, and micrometeoroid impacts. Scaling up to robotic, large-format extrusion systems will be critical for enabling autonomous, ISRU-driven additive manufacturing for long-duration lunar operations.

\section{Conflict of Interest Statement}
On behalf of all authors, the corresponding author states that there is no conflict of interest.

\section{Declaration of Generative AI and AI-assisted technologies in the writing process}
During the preparation of this manuscript, the authors employed Grammarly and ChatGPT to improve language clarity and readability. Subsequently, they reviewed and revised the text as needed and assume full responsibility for its content.

\section{Acknowledgements}

We gratefully acknowledge the financial support of the Natural Sciences and Engineering Research Council of Canada (NSERC) for this project. We also thank Dr.~Marie-Josée Potvin of the Canadian Space Agency (CSA) for her invaluable technical guidance. The authors further acknowledge Farshad Malekpour for his assistance with DSC testing and ideas regarding PEEK-PEKK interfacing, Heng Wang for further DSC testing, and Dr.~Dmytro Kevorkov for his support with microstructural analysis.

%% Loading bibliography style file
% \bibliographystyle{model1-num-names}
\bibliographystyle{unsrt}

% Loading bibliography database
\bibliography{cas-refs}

\begin{thebibliography}{10}

\bibitem{gelino2024selection}
Nathan~J Gelino, Jackson~L Smith, Tesia~D Irwin, Thomas~A Lipscomb, Evan~A Bell, David~I Malott, Stephen~J Pfund, Leonel~H Herrera, Caela~G Gomes, Tracy~L Gibson, et~al.
\newblock Selection, production, and properties of regolith polymer composites for lunar construction.
\newblock In {\em 2024 IEEE Aerospace Conference}, pages 1--21. IEEE, 2024.

\bibitem{azami2024comprehensive}
Mohammad Azami, Zahra Kazemi, Sare Moazen, Martine Dub{\'e}, Marie-Jos{\'e}e Potvin, and Krzysztof Skonieczny.
\newblock A comprehensive review of lunar-based manufacturing and construction.
\newblock {\em Progress in Aerospace Sciences}, 150:101045, 2024.

\bibitem{isachenkov2021regolith}
Maxim Isachenkov, Svyatoslav Chugunov, Iskander Akhatov, and Igor Shishkovsky.
\newblock Regolith-based additive manufacturing for sustainable development of lunar infrastructure--an overview.
\newblock {\em Acta Astronautica}, 180:650--678, 2021.

\bibitem{zocca2022challenges}
Andrea Zocca, Janka Wilbig, Anja Waske, Jens G{\"u}nster, Martinus~Putra Widjaja, Christian Neumann, M{\'e}lanie Clozel, Andreas Meyer, Jifeng Ding, Zuoxin Zhou, et~al.
\newblock Challenges in the technology development for additive manufacturing in space.
\newblock {\em Chinese Journal of Mechanical Engineering: Additive Manufacturing Frontiers}, 1(1):100018, 2022.

\bibitem{altun2021additive}
Altan~Alpay Altun, Florian Ertl, Maude Marechal, Advenit Makaya, Antonella Sgambati, and Martin Schwentenwein.
\newblock Additive manufacturing of lunar regolith structures.
\newblock {\em Open Ceramics}, 5:100058, 2021.

\bibitem{araghi2022nasa}
Koorosh~R Araghi.
\newblock Nasa lunar in-situ resource utilization technology overview.
\newblock In {\em Korean Institute of Geoscience and Mineral Resources (KIGAM) ISRU Workshop}, 2022.

\bibitem{sanders2025progress}
Gerald~Jerry Sanders and Julie Kleinhenz.
\newblock Progress review of nasa lunar isru development: 2019 to 2025.
\newblock In {\em Luxembourg Space Resources Week}. European Space Research and Technology Centre, 2025.

\bibitem{azami2023laser}
Mohammad Azami, Armin Siahsarani, Amir Hadian, Zahra Kazemi, Davood Rahmatabadi, Seyed~Farshid Kashani-Bozorg, and Karen Abrinia.
\newblock Laser powder bed fusion of alumina/fe--ni ceramic matrix particulate composites impregnated with a polymeric resin.
\newblock {\em Journal of Materials Research and Technology}, 24:3133--3144, 2023.

\bibitem{abedi2019high}
HR~Abedi, A~Zarei Hanzaki, M~Azami, M~Kahnooji, and D~Rahmatabadi.
\newblock The high temperature flow behavior of additively manufactured inconel 625 superalloy.
\newblock {\em Materials Research Express}, 6(11):116514, 2019.

\bibitem{kazemi2025uncertainty}
Zahra Kazemi and Craig~A Steeves.
\newblock Uncertainty quantification of local elastic properties in additively manufactured materials for topology optimization applications using machine learning.
\newblock {\em Acta Mechanica}, pages 1--10, 2025.

\bibitem{kazemi2022uncertainty}
Zahra Kazemi and Craig~A Steeves.
\newblock Uncertainty quantification in material properties of additively manufactured materials for application in topology optimization.
\newblock In {\em ASME International Mechanical Engineering Congress and Exposition}, volume 86656, page V003T04A009. American Society of Mechanical Engineers, 2022.

\bibitem{hoffmann2023space}
Miguel Hoffmann and Alaa Elwany.
\newblock In-space additive manufacturing: A review.
\newblock {\em Journal of Manufacturing Science and Engineering}, 145(2):020801, 2023.

\bibitem{li2020slaam}
Jinbo Li, Pierre-Lucas Aubin-Fournier, and Krzysztof Skonieczny.
\newblock Slaam: Simultaneous localization and additive manufacturing.
\newblock {\em IEEE Transactions on Robotics}, 37(2):334--349, 2020.

\bibitem{li2022preparation}
Ruilin Li, Guoqing Zhou, Kang Yan, Jun Chen, Daqing Chen, Shangyue Cai, and Pin-Qiang Mo.
\newblock Preparation and characterization of a specialized lunar regolith simulant for use in lunar low gravity simulation.
\newblock {\em International Journal of Mining Science and Technology}, 32(1):1--15, 2022.

\bibitem{mueller2014additive}
Robert~P Mueller, Laurent Sibille, Paul~E Hintze, Thomas~C Lippitt, James~G Mantovani, Matthew~W Nugent, and Ivan~I Townsend.
\newblock Additive construction using basalt regolith fines.
\newblock In {\em Earth and Space 2014}, pages 394--403. 2014.

\bibitem{creedon2023development}
MJ~Creedon, Tara Linneman, Douglas~L Rickman, and Michael Effinger.
\newblock Development of new lunar highland regolith simulant, nuw-lht-5m.
\newblock In {\em Lunar Surface Innovation Consortium Spring Meeting}, 2023.

\bibitem{wang2022situ}
Yushen Wang, Liang Hao, Yan Li, Qinglei Sun, Mingxi Sun, Yuhong Huang, Zheng Li, Danna Tang, Yijing Wang, and Long Xiao.
\newblock In-situ utilization of regolith resource and future exploration of additive manufacturing for lunar/martian habitats: A review.
\newblock {\em Applied Clay Science}, 229:106673, 2022.

\bibitem{taylor2016evaluations}
Lawrence~A Taylor, Carle~M Pieters, and Daniel Britt.
\newblock Evaluations of lunar regolith simulants.
\newblock {\em Planetary and Space Science}, 126:1--7, 2016.

\bibitem{iantaffi2025laser}
Caterina Iantaffi, Chu Lun~Alex Leung, George Maddison, Eral Bele, Samy Hocine, Rob Snell, Alexander Rack, Martina Meisnar, Thomas Rohr, Iain Todd, et~al.
\newblock Laser additive manufacturing of lunar regolith simulant: New insights from in situ synchrotron x-ray imaging.
\newblock {\em Additive Manufacturing}, 101:104711, 2025.

\bibitem{heiken1991lunar}
Grant Heiken, David Vaniman, and Bevan~M French.
\newblock {\em Lunar sourcebook: A user's guide to the Moon}.
\newblock Number 1259. Cup Archive, 1991.

\bibitem{azami2025enhancing}
Mohammad Azami, Pierre-Lucas Aubin-Fournier, and Krzysztof Skonieczny.
\newblock Enhancing economical lunar-based manufacturing by incorporating lunar regolith into polyether--ether--ketone (peek): material development, additive manufacturing, and characterization.
\newblock {\em Progress in Additive Manufacturing}, pages 1--7, 2025.

\bibitem{prater2016summary}
TJ~Prater, QA~Bean, RD~Beshears, TD~Rolin, NJ~Werkheiser, EA~Ordonez, RM~Ryan, and FE~Ledbetter~III.
\newblock Summary report on phase i results from the 3d printing in zero g technology demonstration mission, volume i.
\newblock Technical report, 2016.

\bibitem{werkheiser2015space}
Niki Werkheiser.
\newblock In-space manufacturing: pioneering a sustainable path to mars.
\newblock Technical report, 2015.

\bibitem{sacco2019additive}
Enea Sacco and Seung~Ki Moon.
\newblock Additive manufacturing for space: status and promises.
\newblock {\em The International Journal of Advanced Manufacturing Technology}, 105:4123--4146, 2019.

\bibitem{huang2025experimental}
Angela Huang and Zheng~H Zhu.
\newblock An experimental investigation of microgravity conditions on fdm-based in-space polymer additive manufacturing.
\newblock {\em Acta Astronautica}, 228:886--899, 2025.

\bibitem{moore1990low}
James~S Moore~Jr, David~R Mengers, Theodore~D Swanson, Reinhard Radermacher, and Frederick~A Costello.
\newblock Low-temperature thermal control for a lunar base.
\newblock {\em SAE Transactions}, pages 598--612, 1990.

\bibitem{marek2016photochemical}
Adam~A Marek and Vincent Verney.
\newblock Photochemical reactivity of pla at the vicinity of glass transition temperature. the photo-rheology method.
\newblock {\em European Polymer Journal}, 81:239--246, 2016.

\bibitem{srithep2013effects}
Yottha Srithep, Paul Nealey, and Lih-Sheng Turng.
\newblock Effects of annealing time and temperature on the crystallinity and heat resistance behavior of injection-molded poly (lactic acid).
\newblock {\em Polymer Engineering \& Science}, 53(3):580--588, 2013.

\bibitem{badr2000mechanism}
Y~Badr, ZI~Ali, and Rasha~M Khafagy.
\newblock On the mechanism of low temperature glass transition in low density polyethylene films.
\newblock {\em Radiation Physics and Chemistry}, 58(1):87--100, 2000.

\bibitem{abbott2017thermoplastic}
Andrew~P Abbott, Tariq~Z Abolibda, Wanwan Qu, William~R Wise, and Luka~A Wright.
\newblock Thermoplastic starch--polyethylene blends homogenised using deep eutectic solvents.
\newblock {\em RSC Advances}, 7(12):7268--7273, 2017.

\bibitem{stehling1970glass}
Ferdinand~C Stehling and Leo Mandelkern.
\newblock The glass temperature of linear polyethylene.
\newblock {\em Macromolecules}, 3(2):242--252, 1970.

\bibitem{zanjanijam2020fused}
Ali~Reza Zanjanijam, Ian Major, John~G Lyons, Ugo Lafont, and Declan~M Devine.
\newblock Fused filament fabrication of peek: A review of process-structure-property relationships.
\newblock {\em Polymers}, 12(8):1665, 2020.

\bibitem{walter1997outgassing}
Neil~A Walter and John~J Scialdone.
\newblock Outgassing data for selecting spacecraft materials.
\newblock Technical report, 1997.

\bibitem{chiggiato2006outgassing}
Paolo Chiggiato.
\newblock Outgassing.
\newblock In {\em CERN Accelerator School}, 2006.

\bibitem{william1993outgassing}
CAMPBELL William.
\newblock Outgassing data for selecting spacecraft materials.
\newblock {\em NASA Reference Publication 1124}, 1993.

\bibitem{pulipaka2023effect}
Aditya Pulipaka, Kunal~Manoj Gide, Ali Beheshti, and Z~Shaghayegh Bagheri.
\newblock Effect of 3d printing process parameters on surface and mechanical properties of fff-printed peek.
\newblock {\em Journal of Manufacturing Processes}, 85:368--386, 2023.

\bibitem{chuang2015additive}
Kathy~C Chuang, Joseph~E Grady, Robert~D Draper, Euy-Sik~E Shin, Clark Patterson, and Thomas~D Santelle.
\newblock Additive manufacturing and characterization of ultem polymers and composites.
\newblock Technical report, 2015.

\bibitem{mclauchlin2014studies}
AR~McLauchlin, OR~Ghita, and L~Savage.
\newblock Studies on the reprocessability of poly (ether ether ketone)(peek).
\newblock {\em Journal of Materials Processing Technology}, 214(1):75--80, 2014.

\bibitem{pascual2019stability}
Alfons Pascual, Michael Toma, Panayota Tsotra, and Markus~C Grob.
\newblock On the stability of peek for short processing cycles at high temperatures and oxygen-containing atmosphere.
\newblock {\em Polymer Degradation and Stability}, 165:161--169, 2019.

\bibitem{azami2024additive}
Mohammad Azami, Pierre-Lucas Aubin~Fournier, and Krzysztof Skonieczny.
\newblock Additive manufacturing of polyether ether ketone (peek)/lunar regolith composites via fused filament fabrication.
\newblock In {\em Earth and Space 2024: Engineering for Extreme Environments}, pages 976--986. 2024.

\bibitem{exolithlms1d}
Exolith.
\newblock Lunar mare dust (lms-1d) moon dirt simulant - fact sheet.
\newblock Accessed on 2023/10/19.

\bibitem{nakamura1992effect}
Yoshinobu Nakamura, Miho Yamaguchi, Masayoshi Okubo, and Tsunetaka Matsumoto.
\newblock Effect of particle size on mechanical properties of epoxy resin filled with angular-shaped silica.
\newblock {\em Journal of applied polymer science}, 44(1):151--158, 1992.

\bibitem{yap2023additive}
Timothy Yap, Nathaniel Heathman, Tim Phillips, Joseph Beaman, and Mehran Tehrani.
\newblock Additive manufacturing of polyaryletherketone (paek) polymers and their composites.
\newblock {\em Composites Part B: Engineering}, 266:111019, 2023.

\bibitem{blundell1983morphology}
David~J Blundell and BN~Osborn.
\newblock The morphology of poly (aryl-ether-ether-ketone).
\newblock {\em Polymer}, 24(8):953--958, 1983.

\bibitem{lai2007peek}
Yen-Huei Lai, MC~Kuo, JC~Huang, and M~Chen.
\newblock On the peek composites reinforced by surface-modified nano-silica.
\newblock {\em Materials Science and Engineering: A}, 458(1-2):158--169, 2007.

\bibitem{arzak1991effect}
A~Arzak, JI~Eguiazabal, and J~Nazabal.
\newblock Effect of annealing on the properties of poly (ether ether ketone).
\newblock {\em Polymer Engineering \& Science}, 31(8):586--591, 1991.

\bibitem{lannunziata2024effect}
Erika Lannunziata, Giovanna Colucci, Paolo Minetola, and Alberto Giubilini.
\newblock Effect of annealing treatment and infill percentage on 3d-printed peek samples by fused filament fabrication.
\newblock {\em The International Journal of Advanced Manufacturing Technology}, 131(9):5209--5222, 2024.

\bibitem{jiang2019effect}
Zhiyuan Jiang, Peng Liu, Hung-Jue Sue, and Tim Bremner.
\newblock Effect of annealing on the viscoelastic behavior of poly (ether-ether-ketone).
\newblock {\em Polymer}, 160:231--237, 2019.

\bibitem{Lau2025_TA}
Hang~Kuen Lau.
\newblock Effect of thermal degradation on polymer thermal properties.
\newblock \url{https://www.tainstruments.com/applications-notes/effect-of-thermal-degradation-on-polymer-thermal-properties/}, 2025.
\newblock TA Instruments Application Note TA430. Controlled chain-scission lowers $T_\mathrm{g}$ by 9\,$^\circ$C. Accessed on May 29, 2025.

\bibitem{sarasua1996effects}
JR~Sarasua, PM~Remiro, and JBAUDB Pouyet.
\newblock Effects of thermal history on mechanical behavior of peek and its short-fiber composites.
\newblock {\em Polymer composites}, 17(3):468--477, 1996.

\bibitem{abraham2009mechanical}
R~Abraham, SP~Thomas, S~Kuryan, J~Isac, KT~Varughese, and S~Thomas.
\newblock Mechanical properties of ceramic-polymer nanocomposites.
\newblock {\em Express Polym Lett}, 3(3):177--89, 2009.

\bibitem{wu2020recent}
Hao Wu, WP~Fahy, Steven Kim, H~Kim, Nanzhu Zhao, Louis Pilato, Abdullah Kafi, Stuart Bateman, and JH~Koo.
\newblock Recent developments in polymers/polymer nanocomposites for additive manufacturing.
\newblock {\em Progress in Materials Science}, 111:100638, 2020.

\bibitem{day1988thermal}
M~Day, T~Suprunchuk, JD~Cooney, and DM~Wiles.
\newblock Thermal degradation of poly (aryl-ether--ether-ketone)(peek): A differential scanning calorimetry study.
\newblock {\em Journal of applied polymer science}, 36(5):1097--1106, 1988.

\bibitem{gaitanelis2023multi}
Dimitrios Gaitanelis, Angeliki Chanteli, Chris Worrall, Paul~M Weaver, and Mihalis Kazilas.
\newblock A multi-technique and multi-scale analysis of the thermal degradation of peek in laser heating.
\newblock {\em Polymer Degradation and Stability}, 211:110282, 2023.

\end{thebibliography}

\end{document}